\documentclass[reprint,aps,prl,twocolumns,superscriptaddress,floatfix,notitlepage]{revtex4-1}
\usepackage[a4paper]{geometry}
\usepackage{graphicx}
\usepackage{float}
\usepackage{amsmath}
\usepackage{amsfonts}
\usepackage{amssymb}
\usepackage{enumitem}
\usepackage{epstopdf}
\usepackage{mathrsfs}
\usepackage{mathtools}
\usepackage{fullpage}
\usepackage[utf8]{inputenc}
\usepackage{xcolor}
\usepackage{bbold}
\usepackage{physics}
\usepackage{bm}
\usepackage[normalem]{ulem}
\def\v{\mathbf{v}}

\def\>{\rangle}
\def\<{\langle}

\newcommand{\be}{\begin{equation}}
\newcommand{\ee}{\end{equation}}
\newcommand{\bea}{\begin{eqnarray}}
\newcommand{\eea}{\end{eqnarray}}

\newcommand{\I} {\mbox{Im}\,}

\renewcommand{\vec}[1]{\mathbf{#1}}
\newcommand*\diff{\mathop{}\!\mathrm{d}}
\usepackage{braket}
\usepackage[colorlinks,bookmarks=true,citecolor=blue,linkcolor=red,urlcolor=blue]{hyperref}
\usepackage{hyperref}

\begin{document}

    \title{$s$-wave paired composite-fermion electron-hole trial state for quantum Hall bilayers with $\nu=1$}
	\author{Glenn Wagner}
	\affiliation{Rudolf Peierls Centre for Theoretical Physics, Univeristy of Oxford, OX1 3PU, UK}
	\affiliation{Kavli Institute for Theoretical Physics, University of California Santa Barbara, CA 93106, USA}
	\author{Dung X. Nguyen}
    \affiliation{Brown Theoretical Physics Center,
Brown University, Providence, RI 02912, USA}
	\author{Steven H. Simon}
	\affiliation{Rudolf Peierls Centre for Theoretical Physics, Univeristy of Oxford, OX1 3PU, UK}
	\author{Bertrand I. Halperin}
	\affiliation{Department  of  Physics,  Harvard  University,  Cambridge,  MA  02138,  USA}

	\begin{abstract}
We introduce a new variational wavefunction for a quantum Hall bilayer at total filling $\nu = 1$, which is based on $s$-wave BCS pairing between composite-fermion electrons in one layer and composite-fermion holes in the other. We compute the overlap of the optimized trial function with the ground state from exact diagonalization calculations of up to 14 electrons in a spherical geometry, and we find excellent agreement over the entire range of values of the ratio between the layer separation and the magnetic length. The trial wavefunction naturally allows for charge imbalance between the layers and provides important insights into how the physics at large interlayer separations crosses over to that at small separations in a fashion analogous to the BEC-BCS crossover.
\end{abstract}

\maketitle	
	
{\it \textbf{ Introduction --}} The bilayer quantum Hall (BQH) system at total filling fraction $\nu=1$
has been the subject of considerable theoretical and experimental interest for more than two decades\cite{Experiment_Review,Eisenstein2004}. The $\nu=1$ BQH system consists of 
two two-dimensional electron systems separated by a distance $d$.   A magnetic field $B$ perpendicular to the layers is  applied such that 
$\nu= n \phi_0/B =1$ with $n = n_\uparrow + n_\downarrow$ the total electron density, and $\phi_0 = 2 \pi \hbar/e$ the flux quantum  ($\uparrow$ and $\downarrow$ refer to the two different layers). 
The competition between inter- and intra-layer Coulomb interactions 
makes this system both interesting and challenging.

 We assume here that all electrons are confined to the lowest Landau level and are fully spin polarized. At small interlayer distances $d$ compared to the magnetic length $\ell_B = \sqrt{\hbar/e B}$ the ground state can be described as an exciton condensate: the electrons in one layer form tightly bound states with the holes in the opposite layer\cite{Experiment_Review,Halperin111}. In contrast, at infinite interlayer distances, the two layers decouple completely, such that each layer forms an independent composite fermion (CF) liquid\cite{HLR,CompositeFermionsJain,CompositeFermionsHeinonen}.   Since these two limits are described in terms of different quasiparticles, understanding how they are connected is a difficult problem.    There has been an enormous amount of theoretical work attempting to address this question\cite{Moon_Review,Eisenstein2004,Bonesteel,p_wave,Pwave2,Ezawa_2009,ShouCheng,HF1,HF2,HF3,HF4,HF5,HF6,ED1,ED3,ED0,DMRG,Park1,Park2,Simon1,Simon2,Simon3,Kimchi,Sodemann,Milovanovic,Ye1,Ye2,ICCFL,Cipri_thesis,CipriBonesteel,papicThesis,Bosonization1,Bosonization4}.

In a recent paper by one of the current authors and one with other collaborators\cite{Halperin2020,Crossover},
a new approach to this crossover was proposed ---  $s$-wave BCS pairing of CFs in one layer with CF-holes (or ``anti-CFs") in the other layer.   This approach qualitatively appeals in that it describes the correct types of quasiparticles both for small $d$ (excitons) and large $d$ (CFs).  Further, it naturally allows a description in the case of imbalanced layers where $n_\uparrow \neq n_\downarrow$. The purpose of this current paper is to numerically test this proposal.


At small $d/\ell_B$, the $\nu=1$ system forms Halperin's (111)-state\cite{Halperin111}.    This state can be viewed as a condensate of interlayer excitons, or equivalently $s$-wave pairing of electrons in one layer with holes in the other. 
This limit is well described in Hartree-Fock\cite{Moon_Review} and is a good description even when the density is imbalanced between the layers.

At large $d/\ell_B$  the description of the layers is more complicated.    For infinite $d/\ell_B$ the layers behave as independent quantum Hall states.   For the balanced case of $\nu_\uparrow = \nu_\downarrow=1/2$  each layer is well described as a CF Fermi sea in zero effective magnetic field\cite{CompositeFermionsJain,CompositeFermionsHeinonen,HLR}.   Away from filling 1/2 the CFs see a residual magnetic field.

When the two $\nu=1/2$ layers are then weakly coupled together, 
we expect the CF liquids to become correlated with each other.  A possibility that was considered from very early on is that the two layers form a BCS paired state of CFs\cite{Bonesteel,Pwave2,p_wave}.   Numerical work with trial wavefunctions established\cite{Simon2} that the $p$-wave channel is the symmetry channel with the largest gap.
In exact diagonalization (ED) studies the $p$-wave paired state was shown\cite{Simon1,Simon2} to have very high overlaps with the exact ground state for $d\gtrsim\ell_B$. In those studies the overlap rapidly decreased at $d\lesssim\ell_B$. However Ref.~\onlinecite{Sodemann} 
argued that the $p$-wave pairing state could be continuously deformed to the exciton condensate without going through a phase transition.

In recent years, after the initial investigations into the $\nu=1$ bilayer, there has been renewed focus on the issue of particle-hole symmetry in the $\nu=1/2$ CF Fermi liquid state\cite{Son,Dirac-Full}.   While the single-layer half-filled Landau level is particle-hole symmetric\cite{DMRG_half_filled}, the CF construction\cite{HLR,CompositeFermionsJain} does not appear to respect this symmetry in any obvious way.    This then raises the question as to whether we should view the half filled Landau level as a Fermi sea of  CF electrons, or as a Fermi sea of composite fermions of {\it holes} removed from a filled Landau level (we call this an ``anti-CF" Fermi sea).    While the two descriptions are numerically almost equivalent\cite{RezayiHaldane}, there may, nonetheless, be advantages to thinking in terms of one or the other.

In this paper we examine a new model of the $\nu=1$ bilayer:  $s$-wave pairing of the CFs in one layer with composite fermions of holes (anti-CFs) in the other layer. 
We show that a trial wavefunction based on this approach has very high overlaps with the exact ground state at all distances $d/\ell_B$.  The evolution of the system as a function of $d/\ell_B$ is analogous to the BEC-BCS crossover familiar from cold atom gases\cite{BCS_BEC}. At large $d$, we have weakly bound CF/anti-CF pairs (BCS limit), whereas at small $d$, we have tightly bound CF/anti-CF pairs,  tending toward the BEC regime. A nice feature of this approach is that if one considers the Chern-Simons (Halperin-Lee-Read\cite{HLR}) description of CFs the  $s$-wave pairing wavefunction described above exactly recovers the $d\rightarrow 0$ limit as the limit where the $s$-wave pairs have very small binding radius and Landau  level mixing is neglected\cite{Crossover,CHEN20051}. 

This approach also applies just as well to the case of charge imbalance of the layers.   If one transfers charge between the layers, the two Fermi seas remain the same size as each other, both growing or shrinking together, so that the pairing is not destroyed.    We find that the imbalanced bilayer system still forms an $s$-wave paired state of CFs and anti-CFs.
 As noted by Yang\cite{KunYang}, the true BEC limit can only be reached in the case of extreme layer imbalance, where the distance between excitons can be much larger than the exciton size.

{\it \textbf{Details of Calculation --}}   We consider Coulomb interaction $e^2/\epsilon r$ for electrons in the same layer (intralayer interaction) and $e^2/\epsilon\sqrt{r^2+d^2}$ for electrons in different layers (interlayer interaction).   As mentioned above,  we assume that the physical spin of the electrons is completely polarized due to the Zeeman splitting and exchange interaction.  We assume zero temperature and no disorder, and we neglect Landau level mixing.  We also assume no tunneling between the layers, which is a good approximation of many of the experiments.

We perform exact diagonalization (ED) on the sphere for systems of up to $N=14$ electrons, 
 i.e. $N_\uparrow= N_\downarrow = 7 $ electrons per layer in the balanced case. 
More generally, the total number of electrons needs to satisfy $N=N_{\uparrow}+N_{\downarrow}=N_\phi+1$ with $N_\phi$ the total number of flux quanta passing through the sphere, such that we have total filling $\nu=1$. We particle-hole transform\cite{PH} the bottom layer, such that we are describing it in terms of $N_{\uparrow}$ hole coordinates.
Note in particular that the number of holes in the bottom layer matches the number of electrons in the top layer.    

In the top layer, we form CFs by attaching two Jastrow factors to each electron. In the planar geometry, this would be achieved by multiplying our wavefunction by $\prod_{i<j}(z_i-z_j)^2$, where $z_i$ is the position of the $i$-th electron in complex notation. (Adaptation to the spherical geometry is discussed in the Supplementary Material\cite{Supplement}.)  The effective flux seen by the CFs is $N_\phi^\textrm{eff} = N_\phi - 2(N_\uparrow-1)$.   In the bottom layer, we similarly form anti-CFs, which see the same effective flux, by attaching two anti-Jastrow factors, multiplying by $\prod_{i<j}(w_i-w_j)^{*2}$, where $w_i$ are the hole coordinates.  We then BCS pair the CFs from the top layer with the anti-CFs from the bottom layer in the $s$-wave channel. We can write down a variational pairing wavefunction based on this approach (see Supplementary Material\cite{Supplement} for details):
\begin{eqnarray}
    \Psi_{\textrm{BCS},s} &=& \prod_{i<j}(z_i-z_j)^2(w_i-w_j)^{*2}\det(G)  \nonumber \\
    G(z_i, w_j)  &=& \sum_{l,m} g_l \,  \phi_{l,m}(z_i)  \phi^*_{l,m}(w_j)  \label{eq:trialwfmaintext}
\end{eqnarray}
where $\phi_{l,m}$ are the Jain-Kamilla orbitals\cite{Kamilla_Jain,Kamilla_thesis} with angular momentum quantum numbers $l,m$  describing CFs in effective flux $N_\phi^\textrm{eff}$.  The $g_l$ are variational parameters.  Due to rotational symmetry, the variational parameters cannot depend on $m$.  This trial wavefunction approach is similar to the BCS $p$-wave pairing approach of Refs.~\cite{Simon1,Simon2}  except that in that work CFs are paired with CFs, whereas here CFs are paired with anti-CFs. 

{In the absence of tunneling, both the total number of electrons and the electron imbalance are conserved separately, giving us a $U_+(1)\times U_-(1)$ symmetry. For a planar geometry let $c_{\vec k\sigma}^\dagger$ create a CF of momentum $\vec k$ in layer $\sigma$ and let $d_{\bf k \sigma}$ create an anti-CF of momentum $\vec k$ in layer $\sigma$.  (Since the CFs see zero effective magnetic field at $\nu=1$, $\bf{k}$ is a good quantum number in this case.) The $p$-wave state has an order parameter $\langle c_{\vec k\uparrow}^\dagger c_{-\vec k\downarrow}^\dagger\rangle$ which breaks $U_+(1)$, whereas the $s$-wave state has an order parameter $\langle c_{\vec k\uparrow}^\dagger d_{\vec k\downarrow}\rangle$ which breaks $U_-(1)$. The order parameter for the Halperin 111 state is $\langle \psi^\dagger_\uparrow(\bf r) \psi_\downarrow(\bf r)\rangle$ where $\psi^\dagger$ is the {\it electron} creation operator.  This also breaks $U_-(1)$ but differs from the $s$-wave order parameter in not having the same Jastrow factors attached.   We note that the proposed order parameter for the  Interlayer Coherent Composite Fermi Liquid (ICCFL) state of   Ref.~\cite{ICCFL} is $\langle c_{\vec k\uparrow}^\dagger c_{\vec k\downarrow}\rangle$, which also breaks $U_-(1)$. But the state attaches Jastrow factors differently from our $s$-wave wavefunction and has poor overlap with exact diagonalization for all values of $d$ (See Supplementary Material\cite{Supplement}).}

We convert the ED ground state into position space and compute the overlap with the trial state by performing Monte-Carlo integration. We use the probability distribution of the (111)-state for the importance sampling. The optimal variational parameters that maximize the overlap are found using a dual annealing algorithm\cite{DA}. 

{\it \textbf{ Balanced case --}}  Let us focus on the case where the two layers are balanced, i.e. $\nu_\uparrow=\nu_\downarrow=1/2$, and the number of electrons per layer on the sphere is $N_{\uparrow}$. 
We show the overlaps of this variational state with the ED ground state in Fig.~\ref{fig:overlaps}a for $N_{\uparrow}=6$ (See Supplementary Material\cite{Supplement} for other system sizes.) For $N_{\uparrow}=4,5,6,7$ we achieve overlaps squared of better than 0.95 by including $3,4,5,6$ variational parameters respectively. Given that the Hilbert space dimensions of the $L^2=0$ subspace in which the ground state and the trial wavefunctions lie are $D(L^2=0)=12,38,252,1599$ respectively, the high overlaps obtained are significant. In  Fig.~\ref{fig:overlaps}a,  we also show overlap results for the (111)-state and the $p$-wave paired state from Ref.~\cite{Simon2}, as well as the CF Fermi liquid state (two uncoupled CF liquids).  

\begin{figure}
    \centering
    \includegraphics[width=\columnwidth]{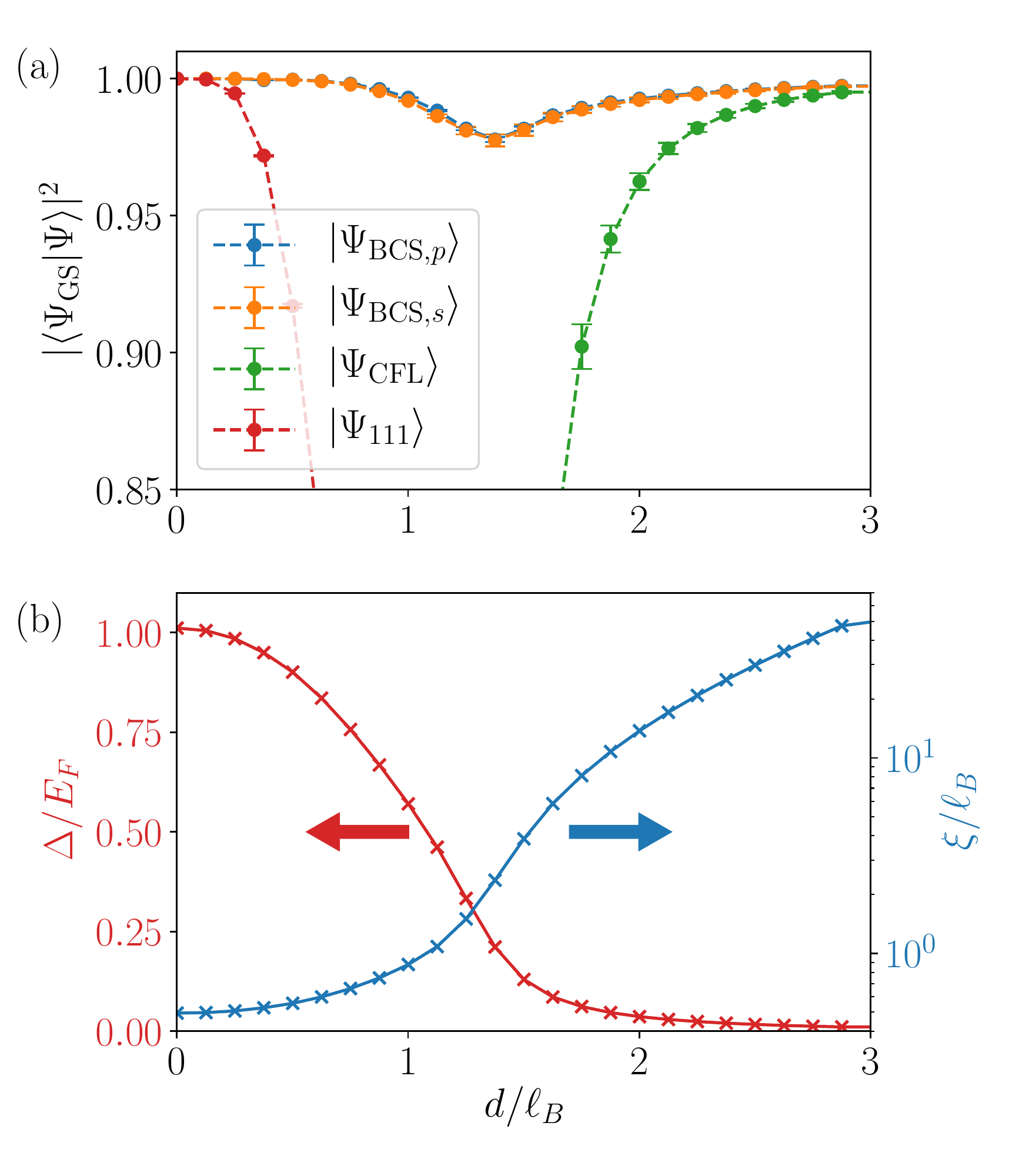}
    \caption{Exact diagonalization results for a balanced system with $6+6$ electrons on the sphere. (a) We plot the overlap of the trial wavefunctions with the true ground state $|\Psi_\textrm{GS}\rangle$ as a function of the interlayer distance $d$. We compare the overlap of our $s$-wave BCS state with the previously proposed $p$-wave BCS state of Refs.~\cite{Simon1,Simon2}. The $s$- and $p$-wave curves are almost indistinguishable on this  plot. For both trial wavefunctions we include 5 variational parameters. We also show the overlaps with the composite Fermi liquid (CFL) state and the 111 state, which are accurate descriptions of the state in the large and small $d$ limits respectively. The errorbars denote the errors of the Monte-Carlo integration. (b) BCS parameters $\Delta/E_F$ and $\xi/\ell_B$ extracted from the $s$-wave variational wavefunction from (a). $\Delta$ is the $s$-wave superconducting  order parameter, $E_F$ is the Fermi energy and $\xi$ is the coherence length. The evolution of the BCS parameters as a function of the interlayer separation $d$ is consistent with a BEC-BCS crossover.}
    \label{fig:overlaps}
\end{figure}

At very large distances the CF Fermi liquid state is essentially exact for $N_{\uparrow}=2,6,12...$ where we have enough CFs to completely fill an integer number of angular momentum shells.  For unfilled shells we construct a Hund's rule state of CFs in each layer where we fill orbitals so as to maximize the angular momentum of each layer\cite{Hund}.    Both the $p$-wave and the $s$-wave variational wavefunctions recover the CF Fermi liquid or Hund's-rule state for a suitable choice of variational parameters, at least when $N$ is such that  one has  a configuration with either an integral number of completely filled angular-momentum shells, or one with a single CF above the outermost filled shell or with a single CF missing from the outermost shell.
At intermediate distances $d/\ell_B \sim 1$ the overlaps of both the $p$ and $s$-wave variational wavefunctions have dips, however they remain extremely accurate in this regime.  At small interlayer distances, the (111)-state is the exact ground state as expected. Both the $p$-wave and the $s$-wave capture this limit as well. As the number of variational parameters is increased, both the $s$-wave and $p$-wave overlaps rapidly improve at small $d$. (In the $s$-wave picture including more variational parameters allows us to form more tightly bound excitons, hence recovering the (111)-state.) Our $s$-wave trial wavefunctions outperform the previous $p$-wave trial state for an equal number of variational parameters.  For example, in Fig.~\ref{fig:overlaps}a, both the $p$ and $s$-wave wavefunctions have five variational parameters.  The squared overlaps at $d=0$ are $0.67$,  $0.95$ and $1.00$ for three, four, and five variational parameters in the $s$-wave case, whereas they are $0.56$, $0.93$ and $1.00$ in the $p$-wave case.  (See Supplementary Material\cite{Supplement} for more details.) 

Note that the $p$-wave wavefunctions described here are putatively the same as those Ref.~\onlinecite{Simon2}.  However, detailed comparison will show that the overlaps with exact diagonalization we obtain here are somewhat better, particularly at small $d$, given the same number (or even fewer) variational parameters. In the present work we use a global optimization algorithm (dual annealing\cite{DA}) to optimize the overlaps. Ref.~\onlinecite{Simon2} used a gradient descent algorithm, which may only find a local optimum of the overlap.

 In Fig.~\ref{fig:overlaps}b we use the best variational $s$-wave trial wavefunction to extract the BCS parameters $\Delta/E_F$ and $\xi$  (see Supplement\cite{Supplement}), where $\xi$ is the coherence length (typical size of a Cooper pair), $\Delta$ is the superconducting order parameter and $E_F$ is the composite fermion Fermi energy.  Note that in regular superconductors $\Delta$ is precisely the excitation gap, but here the superconductor has been ``composite fermionized", so the excitation gap may not precisely match $\Delta$.   We find a crossover from the BEC-like regime ($\xi\lesssim\ell_B$, $\Delta\gtrsim E_F$) at $d\lesssim\ell_B$ to the BCS regime ($\xi\gg\ell_B$, $\Delta\ll E_F$) at $d\gg\ell_B$. In this picture, we have a continuous crossover from the exciton condensate of the 111 state to the BCS-paired composite Fermi liquid.

\begin{figure}
    \centering
    \includegraphics[width=\columnwidth]{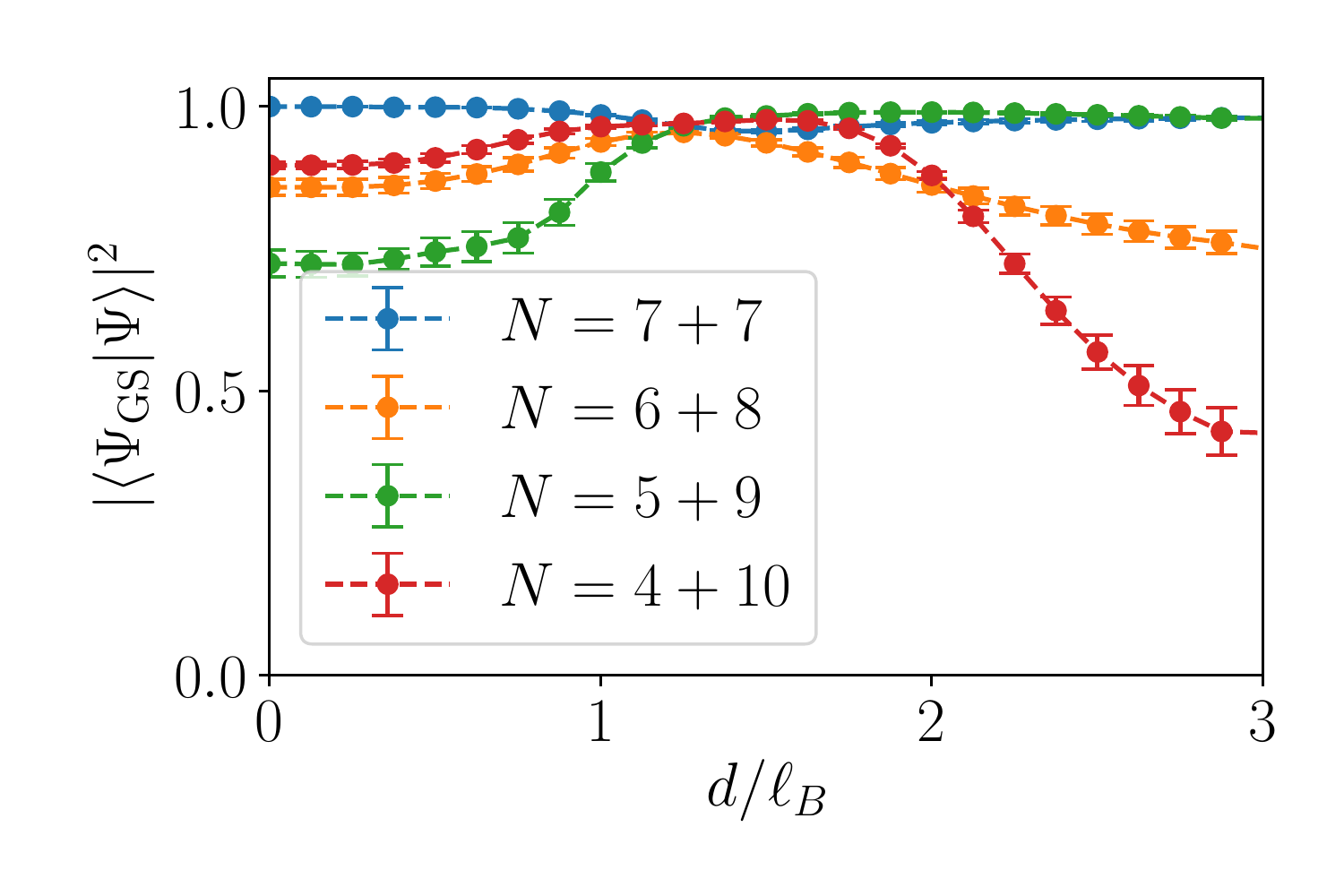}
    \caption{Exact diagonalization results for the overlap of the $s$-wave BCS trial wavefunctions with the true ground state for imbalanced layers with a total of 14 electrons (using 6 variational parameters).  The $L^2=0$ Hilbert space dimensions are 1599, 1319, 614, 205 for $7+7$, $6+8$, $5+9$, and $4+10$ respectively.}
    \label{fig:overlaps_imb}
\end{figure}

{\it \textbf{ Charge imbalance --}} 
We now add a charge imbalance to the two layers, while keeping the total filling fraction constant. The filling fractions of the individual layers are now $\nu_\uparrow=\frac{1-\Delta\nu}{2}$ and $\nu_\downarrow=\frac{1+\Delta\nu}{2}$.   Here we present results for the charge imbalanced $s$-wave CF/anti-CF pairing trial wavefunctions (see Supplementary Material for a discussion of the $p$-wave trial wavefunction for the imbalanced case\cite{Supplement}).  In our approach, we composite-fermionize the minority carriers in each layer, consistent with the experimental observation that the density of the minority carriers sets the Fermi wavevector away from half-filling\cite{minority_carrier,minority_carrier2}.
We show the overlaps of our trial state with the ED ground state in Fig.~\ref{fig:overlaps_imb}.  For small layer imbalances, our trial wavefunction performs less well at small distances than in the balanced case, however considering the dimension of the Hilbert space, the high overlaps obtained even in the imbalanced case are significant. At large $d/\ell_B$, we expect the layers to form independent layers of CFs,  which will successively fill angular-momentum shells.
As mentioned above, our trial wavefunction Eq.~\eqref{eq:trialwfmaintext} can describe this accurately as long as the shells are either filled, or have a single CF in them, or are one CF short of being filled.   This largely explains why some of the values of $(N_\uparrow, N_\downarrow)$ in Fig.~\ref{fig:overlaps_imb} are very accurate at large $d/\ell_B$ and some are inaccurate in this limit.

Experiments observe enhanced superfluid behavior with layer imbalance\cite{Experiments2,Experiments3,Experiments4}. We can conjecture the following natural explanation for this.  At half-filling the CFs (or anti-CFs) are neutral quasiparticles. Away from half-filling the CFs in the top layer develop charge $e (1 - 2 \nu_\uparrow)=e\Delta\nu$, while the anti-CFs in the bottom layer develop charge $e (1 - 2 \nu_\downarrow)=-e\Delta\nu$.  In the imbalanced case, these two charges can attract to improve the BCS pairing. However, once $\Delta\nu\sim 1/2$, we are close to $\frac{1}{4}+\frac{3}{4}$ and the CF description with two flux quanta attached to each electron/hole should be replaced by a CF description where four flux quanta are attached to each electron/hole. A detailed comparison with experiment\cite{Experiments2,Experiments3,Experiments4} would require examination of the energies of possible competing phases, which is beyond the scope of this work. 


{\it \textbf{ Conclusion --}} 
We proposed a new trial wavefunction for the bilayer quantum Hall system, where CFs and anti-CFs pair up in the $s$-wave channel. This trial state has very high overlaps with the exact ground state for any interlayer separation $d$.  In this language, the bilayer system undergoes a BEC-BCS crossover as the interlayer separation is varied. At large $d$ the system is in the BCS limit, with weakly bound CF/anti-CF Cooper pairs, whereas at small $d$, the system enters the BEC regime with tightly bound CF/anti-CF excitons (equivalent to electron/hole excitons in the tightly bound limit\footnote{Note that the phases of the Jastrow factors $(z_i-z_j)^2(w_i-w_j)^{*2}$ cancel in the tightly bound limit where the CFs and anti-CFs are at the same position.}). Our trial state also performs extremely well for imbalanced layers.

We also re-examined the trial wavefunction based on pairing CFs in both layers in the $p$-wave channel and found that by including sufficiently many variational parameters and by using an improved optimization algorithm compared to Ref.~\cite{Simon1} this wavefunction can also accurately describe the system for any interlayer separation $d$. This is consistent with Ref.~\cite{Sodemann}, which used field theory arguments to show that the $p$-wave state can be continuously connected to the (111)-state. The $p$-wave pairing of Halperin-Lee-Read (HLR) CFs corresponds to $s$-wave pairing of Dirac CFs\cite{Son}, due to the Berry phase of the Dirac CFs.  Ref.~\cite{Sodemann} has  used this picture of $s$-wave pairing of Dirac CFs to study the problem of BQH in the balanced case.  However, as far as we are aware, there are no useful trial wavefunctions based on Dirac CFs for either single or double layer systems.



{\it \textbf{{Acknowledgements} --}} Numerical calculations were performed using the DiagHam library. We thank Nicolas Regnault for assistance with DiagHam. GW would like to thank Gunnar M\"oller, Ajit  Balram and Frank Pollmann for useful discussions. GW thanks the Kavli Institute for Theoretical Physics for its hospitality during the graduate fellowship programme. This research was supported in part by the National Science Foundation under Grant No. NSF PHY-1748958 and by the Heising-Simons Foundation. DXN is supported by the Brown Theoretical Physics Center. SHS is supported from EPSRC Grant EP/S020527/1. Statement  of  compliance  with  EPSRC  policy  framework  on  research  data:  This  publication is theoretical work that does not require supporting research data.  BIH is supported in part by the Science and Technology Center for Integrated Quantum Materials, under NSF grant DMR-1231319.

\bibliography{bib}

\begin{thebibliography}{63}%
\makeatletter
\providecommand \@ifxundefined [1]{%
 \@ifx{#1\undefined}
}%
\providecommand \@ifnum [1]{%
 \ifnum #1\expandafter \@firstoftwo
 \else \expandafter \@secondoftwo
 \fi
}%
\providecommand \@ifx [1]{%
 \ifx #1\expandafter \@firstoftwo
 \else \expandafter \@secondoftwo
 \fi
}%
\providecommand \natexlab [1]{#1}%
\providecommand \enquote  [1]{``#1''}%
\providecommand \bibnamefont  [1]{#1}%
\providecommand \bibfnamefont [1]{#1}%
\providecommand \citenamefont [1]{#1}%
\providecommand \href@noop [0]{\@secondoftwo}%
\providecommand \href [0]{\begingroup \@sanitize@url \@href}%
\providecommand \@href[1]{\@@startlink{#1}\@@href}%
\providecommand \@@href[1]{\endgroup#1\@@endlink}%
\providecommand \@sanitize@url [0]{\catcode `\\12\catcode `\$12\catcode
  `\&12\catcode `\#12\catcode `\^12\catcode `\_12\catcode `\%12\relax}%
\providecommand \@@startlink[1]{}%
\providecommand \@@endlink[0]{}%
\providecommand \url  [0]{\begingroup\@sanitize@url \@url }%
\providecommand \@url [1]{\endgroup\@href {#1}{\urlprefix }}%
\providecommand \urlprefix  [0]{URL }%
\providecommand \Eprint [0]{\href }%
\providecommand \doibase [0]{http://dx.doi.org/}%
\providecommand \selectlanguage [0]{\@gobble}%
\providecommand \bibinfo  [0]{\@secondoftwo}%
\providecommand \bibfield  [0]{\@secondoftwo}%
\providecommand \translation [1]{[#1]}%
\providecommand \BibitemOpen [0]{}%
\providecommand \bibitemStop [0]{}%
\providecommand \bibitemNoStop [0]{.\EOS\space}%
\providecommand \EOS [0]{\spacefactor3000\relax}%
\providecommand \BibitemShut  [1]{\csname bibitem#1\endcsname}%
\let\auto@bib@innerbib\@empty
\bibitem [{\citenamefont {Eisenstein}(2014)}]{Experiment_Review}%
  \BibitemOpen
  \bibfield  {author} {\bibinfo {author} {\bibfnamefont {J.}~\bibnamefont
  {Eisenstein}},\ }\href {\doibase 10.1146/annurev-conmatphys-031113-133832}
  {\bibfield  {journal} {\bibinfo  {journal} {Annual Review of Condensed Matter
  Physics}\ }\textbf {\bibinfo {volume} {5}},\ \bibinfo {pages} {159} (\bibinfo
  {year} {2014})}\BibitemShut {NoStop}%
\bibitem [{\citenamefont {Eisenstein}\ and\ \citenamefont
  {MacDonald}(2004)}]{Eisenstein2004}%
  \BibitemOpen
  \bibfield  {author} {\bibinfo {author} {\bibfnamefont {J.~P.}\ \bibnamefont
  {Eisenstein}}\ and\ \bibinfo {author} {\bibfnamefont {A.~H.}\ \bibnamefont
  {MacDonald}},\ }\href {\doibase 10.1038/nature03081} {\bibfield  {journal}
  {\bibinfo  {journal} {Nature}\ }\textbf {\bibinfo {volume} {432}},\ \bibinfo
  {pages} {691} (\bibinfo {year} {2004})}\BibitemShut {NoStop}%
\bibitem [{\citenamefont {Halperin}(1983)}]{Halperin111}%
  \BibitemOpen
  \bibfield  {author} {\bibinfo {author} {\bibfnamefont {B.~I.}\ \bibnamefont
  {Halperin}},\ }\href@noop {} {\bibfield  {journal} {\bibinfo  {journal}
  {Helv. Phys. Acta}\ }\textbf {\bibinfo {volume} {56}},\ \bibinfo {pages} {75}
  (\bibinfo {year} {1983})}\BibitemShut {NoStop}%
\bibitem [{\citenamefont {Halperin}\ \emph {et~al.}(1993)\citenamefont
  {Halperin}, \citenamefont {Lee},\ and\ \citenamefont {Read}}]{HLR}%
  \BibitemOpen
  \bibfield  {author} {\bibinfo {author} {\bibfnamefont {B.~I.}\ \bibnamefont
  {Halperin}}, \bibinfo {author} {\bibfnamefont {P.~A.}\ \bibnamefont {Lee}}, \
  and\ \bibinfo {author} {\bibfnamefont {N.}~\bibnamefont {Read}},\ }\href
  {\doibase 10.1103/PhysRevB.47.7312} {\bibfield  {journal} {\bibinfo
  {journal} {Phys. Rev. B}\ }\textbf {\bibinfo {volume} {47}},\ \bibinfo
  {pages} {7312} (\bibinfo {year} {1993})}\BibitemShut {NoStop}%
\bibitem [{\citenamefont {Jain}(2007)}]{CompositeFermionsJain}%
  \BibitemOpen
  \bibfield  {author} {\bibinfo {author} {\bibfnamefont {J.~K.}\ \bibnamefont
  {Jain}},\ }\href@noop {} {\emph {\bibinfo {title} {Composite Fermions}}}\
  (\bibinfo  {publisher} {Cambridge University Press},\ \bibinfo {year}
  {2007})\BibitemShut {NoStop}%
\bibitem [{\citenamefont {Heinonen}(1998)}]{CompositeFermionsHeinonen}%
  \BibitemOpen
  \bibinfo {editor} {\bibfnamefont {O.}~\bibnamefont {Heinonen}},\ ed.,\
  \href@noop {} {\emph {\bibinfo {title} {Composite Fermions: A Unified View of
  the Quantum Hall Regime}}}\ (\bibinfo  {publisher} {World Scientific},\
  \bibinfo {year} {1998})\BibitemShut {NoStop}%
\bibitem [{\citenamefont {Moon}\ \emph {et~al.}(1995)\citenamefont {Moon},
  \citenamefont {Mori}, \citenamefont {Yang}, \citenamefont {Girvin},
  \citenamefont {MacDonald}, \citenamefont {Zheng}, \citenamefont {Yoshioka},\
  and\ \citenamefont {Zhang}}]{Moon_Review}%
  \BibitemOpen
  \bibfield  {author} {\bibinfo {author} {\bibfnamefont {K.}~\bibnamefont
  {Moon}}, \bibinfo {author} {\bibfnamefont {H.}~\bibnamefont {Mori}}, \bibinfo
  {author} {\bibfnamefont {K.}~\bibnamefont {Yang}}, \bibinfo {author}
  {\bibfnamefont {S.~M.}\ \bibnamefont {Girvin}}, \bibinfo {author}
  {\bibfnamefont {A.~H.}\ \bibnamefont {MacDonald}}, \bibinfo {author}
  {\bibfnamefont {L.}~\bibnamefont {Zheng}}, \bibinfo {author} {\bibfnamefont
  {D.}~\bibnamefont {Yoshioka}}, \ and\ \bibinfo {author} {\bibfnamefont
  {S.-C.}\ \bibnamefont {Zhang}},\ }\href {\doibase 10.1103/PhysRevB.51.5138}
  {\bibfield  {journal} {\bibinfo  {journal} {Phys. Rev. B}\ }\textbf {\bibinfo
  {volume} {51}},\ \bibinfo {pages} {5138} (\bibinfo {year}
  {1995})}\BibitemShut {NoStop}%
\bibitem [{\citenamefont {Bonesteel}\ \emph {et~al.}(1996)\citenamefont
  {Bonesteel}, \citenamefont {McDonald},\ and\ \citenamefont
  {Nayak}}]{Bonesteel}%
  \BibitemOpen
  \bibfield  {author} {\bibinfo {author} {\bibfnamefont {N.~E.}\ \bibnamefont
  {Bonesteel}}, \bibinfo {author} {\bibfnamefont {I.~A.}\ \bibnamefont
  {McDonald}}, \ and\ \bibinfo {author} {\bibfnamefont {C.}~\bibnamefont
  {Nayak}},\ }\href {\doibase 10.1103/PhysRevLett.77.3009} {\bibfield
  {journal} {\bibinfo  {journal} {Phys. Rev. Lett.}\ }\textbf {\bibinfo
  {volume} {77}},\ \bibinfo {pages} {3009} (\bibinfo {year}
  {1996})}\BibitemShut {NoStop}%
\bibitem [{\citenamefont {Isobe}\ and\ \citenamefont {Fu}(2017)}]{p_wave}%
  \BibitemOpen
  \bibfield  {author} {\bibinfo {author} {\bibfnamefont {H.}~\bibnamefont
  {Isobe}}\ and\ \bibinfo {author} {\bibfnamefont {L.}~\bibnamefont {Fu}},\
  }\href {\doibase 10.1103/PhysRevLett.118.166401} {\bibfield  {journal}
  {\bibinfo  {journal} {Phys. Rev. Lett.}\ }\textbf {\bibinfo {volume} {118}},\
  \bibinfo {pages} {166401} (\bibinfo {year} {2017})}\BibitemShut {NoStop}%
\bibitem [{\citenamefont {Morinari}(1999)}]{Pwave2}%
  \BibitemOpen
  \bibfield  {author} {\bibinfo {author} {\bibfnamefont {T.}~\bibnamefont
  {Morinari}},\ }\href {\doibase 10.1103/PhysRevB.59.7320} {\bibfield
  {journal} {\bibinfo  {journal} {Phys. Rev. B}\ }\textbf {\bibinfo {volume}
  {59}},\ \bibinfo {pages} {7320} (\bibinfo {year} {1999})}\BibitemShut
  {NoStop}%
\bibitem [{\citenamefont {Ezawa}\ and\ \citenamefont
  {Tsitsishvili}(2009)}]{Ezawa_2009}%
  \BibitemOpen
  \bibfield  {author} {\bibinfo {author} {\bibfnamefont {Z.~F.}\ \bibnamefont
  {Ezawa}}\ and\ \bibinfo {author} {\bibfnamefont {G.}~\bibnamefont
  {Tsitsishvili}},\ }\href {\doibase 10.1088/0034-4885/72/8/086502} {\bibfield
  {journal} {\bibinfo  {journal} {Reports on Progress in Physics}\ }\textbf
  {\bibinfo {volume} {72}},\ \bibinfo {pages} {086502} (\bibinfo {year}
  {2009})}\BibitemShut {NoStop}%
\bibitem [{\citenamefont {Lian}\ and\ \citenamefont {Zhang}(2018)}]{ShouCheng}%
  \BibitemOpen
  \bibfield  {author} {\bibinfo {author} {\bibfnamefont {B.}~\bibnamefont
  {Lian}}\ and\ \bibinfo {author} {\bibfnamefont {S.-C.}\ \bibnamefont
  {Zhang}},\ }\href {\doibase 10.1103/PhysRevLett.120.077601} {\bibfield
  {journal} {\bibinfo  {journal} {Phys. Rev. Lett.}\ }\textbf {\bibinfo
  {volume} {120}},\ \bibinfo {pages} {077601} (\bibinfo {year}
  {2018})}\BibitemShut {NoStop}%
\bibitem [{\citenamefont {Joglekar}\ and\ \citenamefont
  {MacDonald}(2001{\natexlab{a}})}]{HF1}%
  \BibitemOpen
  \bibfield  {author} {\bibinfo {author} {\bibfnamefont {Y.~N.}\ \bibnamefont
  {Joglekar}}\ and\ \bibinfo {author} {\bibfnamefont {A.~H.}\ \bibnamefont
  {MacDonald}},\ }\href {\doibase 10.1103/PhysRevB.64.155315} {\bibfield
  {journal} {\bibinfo  {journal} {Phys. Rev. B}\ }\textbf {\bibinfo {volume}
  {64}},\ \bibinfo {pages} {155315} (\bibinfo {year}
  {2001}{\natexlab{a}})}\BibitemShut {NoStop}%
\bibitem [{\citenamefont {Joglekar}\ and\ \citenamefont
  {MacDonald}(2001{\natexlab{b}})}]{HF2}%
  \BibitemOpen
  \bibfield  {author} {\bibinfo {author} {\bibfnamefont {Y.~N.}\ \bibnamefont
  {Joglekar}}\ and\ \bibinfo {author} {\bibfnamefont {A.~H.}\ \bibnamefont
  {MacDonald}},\ }\href {\doibase 10.1103/PhysRevLett.87.196802} {\bibfield
  {journal} {\bibinfo  {journal} {Phys. Rev. Lett.}\ }\textbf {\bibinfo
  {volume} {87}},\ \bibinfo {pages} {196802} (\bibinfo {year}
  {2001}{\natexlab{b}})}\BibitemShut {NoStop}%
\bibitem [{\citenamefont {Joglekar}\ and\ \citenamefont
  {MacDonald}(2002)}]{HF3}%
  \BibitemOpen
  \bibfield  {author} {\bibinfo {author} {\bibfnamefont {Y.~N.}\ \bibnamefont
  {Joglekar}}\ and\ \bibinfo {author} {\bibfnamefont {A.~H.}\ \bibnamefont
  {MacDonald}},\ }\href {\doibase 10.1103/PhysRevB.65.235319} {\bibfield
  {journal} {\bibinfo  {journal} {Phys. Rev. B}\ }\textbf {\bibinfo {volume}
  {65}},\ \bibinfo {pages} {235319} (\bibinfo {year} {2002})}\BibitemShut
  {NoStop}%
\bibitem [{\citenamefont {MacDonald}\ \emph {et~al.}(1990)\citenamefont
  {MacDonald}, \citenamefont {Platzman},\ and\ \citenamefont
  {Boebinger}}]{HF4}%
  \BibitemOpen
  \bibfield  {author} {\bibinfo {author} {\bibfnamefont {A.~H.}\ \bibnamefont
  {MacDonald}}, \bibinfo {author} {\bibfnamefont {P.~M.}\ \bibnamefont
  {Platzman}}, \ and\ \bibinfo {author} {\bibfnamefont {G.~S.}\ \bibnamefont
  {Boebinger}},\ }\href {\doibase 10.1103/PhysRevLett.65.775} {\bibfield
  {journal} {\bibinfo  {journal} {Phys. Rev. Lett.}\ }\textbf {\bibinfo
  {volume} {65}},\ \bibinfo {pages} {775} (\bibinfo {year} {1990})}\BibitemShut
  {NoStop}%
\bibitem [{\citenamefont {Fertig}(1989)}]{HF5}%
  \BibitemOpen
  \bibfield  {author} {\bibinfo {author} {\bibfnamefont {H.~A.}\ \bibnamefont
  {Fertig}},\ }\href {\doibase 10.1103/PhysRevB.40.1087} {\bibfield  {journal}
  {\bibinfo  {journal} {Phys. Rev. B}\ }\textbf {\bibinfo {volume} {40}},\
  \bibinfo {pages} {1087} (\bibinfo {year} {1989})}\BibitemShut {NoStop}%
\bibitem [{\citenamefont {C\^ot\'e}\ \emph {et~al.}(1992)\citenamefont
  {C\^ot\'e}, \citenamefont {Brey},\ and\ \citenamefont {MacDonald}}]{HF6}%
  \BibitemOpen
  \bibfield  {author} {\bibinfo {author} {\bibfnamefont {R.}~\bibnamefont
  {C\^ot\'e}}, \bibinfo {author} {\bibfnamefont {L.}~\bibnamefont {Brey}}, \
  and\ \bibinfo {author} {\bibfnamefont {A.~H.}\ \bibnamefont {MacDonald}},\
  }\href {\doibase 10.1103/PhysRevB.46.10239} {\bibfield  {journal} {\bibinfo
  {journal} {Phys. Rev. B}\ }\textbf {\bibinfo {volume} {46}},\ \bibinfo
  {pages} {10239} (\bibinfo {year} {1992})}\BibitemShut {NoStop}%
\bibitem [{\citenamefont {Zhu}\ \emph {et~al.}(2017)\citenamefont {Zhu},
  \citenamefont {Fu},\ and\ \citenamefont {Sheng}}]{ED1}%
  \BibitemOpen
  \bibfield  {author} {\bibinfo {author} {\bibfnamefont {Z.}~\bibnamefont
  {Zhu}}, \bibinfo {author} {\bibfnamefont {L.}~\bibnamefont {Fu}}, \ and\
  \bibinfo {author} {\bibfnamefont {D.~N.}\ \bibnamefont {Sheng}},\ }\href
  {\doibase 10.1103/PhysRevLett.119.177601} {\bibfield  {journal} {\bibinfo
  {journal} {Phys. Rev. Lett.}\ }\textbf {\bibinfo {volume} {119}},\ \bibinfo
  {pages} {177601} (\bibinfo {year} {2017})}\BibitemShut {NoStop}%
\bibitem [{\citenamefont {Nomura}\ and\ \citenamefont {Yoshioka}(2002)}]{ED3}%
  \BibitemOpen
  \bibfield  {author} {\bibinfo {author} {\bibfnamefont {K.}~\bibnamefont
  {Nomura}}\ and\ \bibinfo {author} {\bibfnamefont {D.}~\bibnamefont
  {Yoshioka}},\ }\href {\doibase 10.1103/PhysRevB.66.153310} {\bibfield
  {journal} {\bibinfo  {journal} {Phys. Rev. B}\ }\textbf {\bibinfo {volume}
  {66}},\ \bibinfo {pages} {153310} (\bibinfo {year} {2002})}\BibitemShut
  {NoStop}%
\bibitem [{\citenamefont {Schliemann}\ \emph {et~al.}(2001)\citenamefont
  {Schliemann}, \citenamefont {Girvin},\ and\ \citenamefont {MacDonald}}]{ED0}%
  \BibitemOpen
  \bibfield  {author} {\bibinfo {author} {\bibfnamefont {J.}~\bibnamefont
  {Schliemann}}, \bibinfo {author} {\bibfnamefont {S.~M.}\ \bibnamefont
  {Girvin}}, \ and\ \bibinfo {author} {\bibfnamefont {A.~H.}\ \bibnamefont
  {MacDonald}},\ }\href {\doibase 10.1103/PhysRevLett.86.1849} {\bibfield
  {journal} {\bibinfo  {journal} {Phys. Rev. Lett.}\ }\textbf {\bibinfo
  {volume} {86}},\ \bibinfo {pages} {1849} (\bibinfo {year}
  {2001})}\BibitemShut {NoStop}%
\bibitem [{\citenamefont {Shibata}\ and\ \citenamefont
  {Yoshioka}(2006)}]{DMRG}%
  \BibitemOpen
  \bibfield  {author} {\bibinfo {author} {\bibfnamefont {N.}~\bibnamefont
  {Shibata}}\ and\ \bibinfo {author} {\bibfnamefont {D.}~\bibnamefont
  {Yoshioka}},\ }\href@noop {} {\bibfield  {journal} {\bibinfo  {journal}
  {Journal of the Physical Society of Japan}\ }\textbf {\bibinfo {volume}
  {75}},\ \bibinfo {pages} {043712} (\bibinfo {year} {2006})}\BibitemShut
  {NoStop}%
\bibitem [{\citenamefont {Park}(2004)}]{Park1}%
  \BibitemOpen
  \bibfield  {author} {\bibinfo {author} {\bibfnamefont {K.}~\bibnamefont
  {Park}},\ }\href {\doibase 10.1103/PhysRevB.69.045319} {\bibfield  {journal}
  {\bibinfo  {journal} {Phys. Rev. B}\ }\textbf {\bibinfo {volume} {69}},\
  \bibinfo {pages} {045319} (\bibinfo {year} {2004})}\BibitemShut {NoStop}%
\bibitem [{\citenamefont {Park}\ and\ \citenamefont {Das~Sarma}(2006)}]{Park2}%
  \BibitemOpen
  \bibfield  {author} {\bibinfo {author} {\bibfnamefont {K.}~\bibnamefont
  {Park}}\ and\ \bibinfo {author} {\bibfnamefont {S.}~\bibnamefont
  {Das~Sarma}},\ }\href {\doibase 10.1103/PhysRevB.74.035338} {\bibfield
  {journal} {\bibinfo  {journal} {Phys. Rev. B}\ }\textbf {\bibinfo {volume}
  {74}},\ \bibinfo {pages} {035338} (\bibinfo {year} {2006})}\BibitemShut
  {NoStop}%
\bibitem [{\citenamefont {M\"oller}\ \emph {et~al.}(2009)\citenamefont
  {M\"oller}, \citenamefont {Simon},\ and\ \citenamefont {Rezayi}}]{Simon1}%
  \BibitemOpen
  \bibfield  {author} {\bibinfo {author} {\bibfnamefont {G.}~\bibnamefont
  {M\"oller}}, \bibinfo {author} {\bibfnamefont {S.~H.}\ \bibnamefont {Simon}},
  \ and\ \bibinfo {author} {\bibfnamefont {E.~H.}\ \bibnamefont {Rezayi}},\
  }\href {\doibase 10.1103/PhysRevB.79.125106} {\bibfield  {journal} {\bibinfo
  {journal} {Phys. Rev. B}\ }\textbf {\bibinfo {volume} {79}},\ \bibinfo
  {pages} {125106} (\bibinfo {year} {2009})}\BibitemShut {NoStop}%
\bibitem [{\citenamefont {M\"oller}\ \emph {et~al.}(2008)\citenamefont
  {M\"oller}, \citenamefont {Simon},\ and\ \citenamefont {Rezayi}}]{Simon2}%
  \BibitemOpen
  \bibfield  {author} {\bibinfo {author} {\bibfnamefont {G.}~\bibnamefont
  {M\"oller}}, \bibinfo {author} {\bibfnamefont {S.~H.}\ \bibnamefont {Simon}},
  \ and\ \bibinfo {author} {\bibfnamefont {E.~H.}\ \bibnamefont {Rezayi}},\
  }\href {\doibase 10.1103/PhysRevLett.101.176803} {\bibfield  {journal}
  {\bibinfo  {journal} {Phys. Rev. Lett.}\ }\textbf {\bibinfo {volume} {101}},\
  \bibinfo {pages} {176803} (\bibinfo {year} {2008})}\BibitemShut {NoStop}%
\bibitem [{\citenamefont {Simon}\ \emph {et~al.}(2003)\citenamefont {Simon},
  \citenamefont {Rezayi},\ and\ \citenamefont {Milovanovic}}]{Simon3}%
  \BibitemOpen
  \bibfield  {author} {\bibinfo {author} {\bibfnamefont {S.~H.}\ \bibnamefont
  {Simon}}, \bibinfo {author} {\bibfnamefont {E.~H.}\ \bibnamefont {Rezayi}}, \
  and\ \bibinfo {author} {\bibfnamefont {M.~V.}\ \bibnamefont {Milovanovic}},\
  }\href {\doibase 10.1103/PhysRevLett.91.046803} {\bibfield  {journal}
  {\bibinfo  {journal} {Phys. Rev. Lett.}\ }\textbf {\bibinfo {volume} {91}},\
  \bibinfo {pages} {046803} (\bibinfo {year} {2003})}\BibitemShut {NoStop}%
\bibitem [{\citenamefont {{Zhang}}\ and\ \citenamefont
  {{Kimchi}}(2018)}]{Kimchi}%
  \BibitemOpen
  \bibfield  {author} {\bibinfo {author} {\bibfnamefont {Y.-H.}\ \bibnamefont
  {{Zhang}}}\ and\ \bibinfo {author} {\bibfnamefont {I.}~\bibnamefont
  {{Kimchi}}},\ }\href@noop {} {\bibfield  {journal} {\bibinfo  {journal}
  {arXiv e-prints}\ ,\ \bibinfo {eid} {arXiv:1810.02809}} (\bibinfo {year}
  {2018})},\ \Eprint {http://arxiv.org/abs/1810.02809} {arXiv:1810.02809
  [cond-mat.str-el]} \BibitemShut {NoStop}%
\bibitem [{\citenamefont {Sodemann}\ \emph {et~al.}(2017)\citenamefont
  {Sodemann}, \citenamefont {Kimchi}, \citenamefont {Wang},\ and\ \citenamefont
  {Senthil}}]{Sodemann}%
  \BibitemOpen
  \bibfield  {author} {\bibinfo {author} {\bibfnamefont {I.}~\bibnamefont
  {Sodemann}}, \bibinfo {author} {\bibfnamefont {I.}~\bibnamefont {Kimchi}},
  \bibinfo {author} {\bibfnamefont {C.}~\bibnamefont {Wang}}, \ and\ \bibinfo
  {author} {\bibfnamefont {T.}~\bibnamefont {Senthil}},\ }\href {\doibase
  10.1103/PhysRevB.95.085135} {\bibfield  {journal} {\bibinfo  {journal} {Phys.
  Rev. B}\ }\textbf {\bibinfo {volume} {95}},\ \bibinfo {pages} {085135}
  (\bibinfo {year} {2017})}\BibitemShut {NoStop}%
\bibitem [{\citenamefont {Milovanovi\ifmmode~\acute{c}\else \'{c}\fi{}}\ \emph
  {et~al.}(2015)\citenamefont {Milovanovi\ifmmode~\acute{c}\else \'{c}\fi{}},
  \citenamefont {Dobard\ifmmode \check{z}\else
  \v{z}\fi{}i\ifmmode~\acute{c}\else \'{c}\fi{}},\ and\ \citenamefont
  {Papi\ifmmode~\acute{c}\else \'{c}\fi{}}}]{Milovanovic}%
  \BibitemOpen
  \bibfield  {author} {\bibinfo {author} {\bibfnamefont {M.~V.}\ \bibnamefont
  {Milovanovi\ifmmode~\acute{c}\else \'{c}\fi{}}}, \bibinfo {author}
  {\bibfnamefont {E.}~\bibnamefont {Dobard\ifmmode \check{z}\else
  \v{z}\fi{}i\ifmmode~\acute{c}\else \'{c}\fi{}}}, \ and\ \bibinfo {author}
  {\bibfnamefont {Z.}~\bibnamefont {Papi\ifmmode~\acute{c}\else \'{c}\fi{}}},\
  }\href {\doibase 10.1103/PhysRevB.92.195311} {\bibfield  {journal} {\bibinfo
  {journal} {Phys. Rev. B}\ }\textbf {\bibinfo {volume} {92}},\ \bibinfo
  {pages} {195311} (\bibinfo {year} {2015})}\BibitemShut {NoStop}%
\bibitem [{\citenamefont {Ye}(2006)}]{Ye1}%
  \BibitemOpen
  \bibfield  {author} {\bibinfo {author} {\bibfnamefont {J.}~\bibnamefont
  {Ye}},\ }\href {\doibase 10.1103/PhysRevLett.97.236803} {\bibfield  {journal}
  {\bibinfo  {journal} {Phys. Rev. Lett.}\ }\textbf {\bibinfo {volume} {97}},\
  \bibinfo {pages} {236803} (\bibinfo {year} {2006})}\BibitemShut {NoStop}%
\bibitem [{\citenamefont {Ye}\ and\ \citenamefont {Jiang}(2007)}]{Ye2}%
  \BibitemOpen
  \bibfield  {author} {\bibinfo {author} {\bibfnamefont {J.}~\bibnamefont
  {Ye}}\ and\ \bibinfo {author} {\bibfnamefont {L.}~\bibnamefont {Jiang}},\
  }\href {\doibase 10.1103/PhysRevLett.98.236802} {\bibfield  {journal}
  {\bibinfo  {journal} {Phys. Rev. Lett.}\ }\textbf {\bibinfo {volume} {98}},\
  \bibinfo {pages} {236802} (\bibinfo {year} {2007})}\BibitemShut {NoStop}%
\bibitem [{\citenamefont {Alicea}\ \emph {et~al.}(2009)\citenamefont {Alicea},
  \citenamefont {Motrunich}, \citenamefont {Refael},\ and\ \citenamefont
  {Fisher}}]{ICCFL}%
  \BibitemOpen
  \bibfield  {author} {\bibinfo {author} {\bibfnamefont {J.}~\bibnamefont
  {Alicea}}, \bibinfo {author} {\bibfnamefont {O.~I.}\ \bibnamefont
  {Motrunich}}, \bibinfo {author} {\bibfnamefont {G.}~\bibnamefont {Refael}}, \
  and\ \bibinfo {author} {\bibfnamefont {M.~P.~A.}\ \bibnamefont {Fisher}},\
  }\href {\doibase 10.1103/PhysRevLett.103.256403} {\bibfield  {journal}
  {\bibinfo  {journal} {Phys. Rev. Lett.}\ }\textbf {\bibinfo {volume} {103}},\
  \bibinfo {pages} {256403} (\bibinfo {year} {2009})}\BibitemShut {NoStop}%
\bibitem [{\citenamefont {Cipri}(2014)}]{Cipri_thesis}%
  \BibitemOpen
  \bibfield  {author} {\bibinfo {author} {\bibfnamefont {R.}~\bibnamefont
  {Cipri}},\ }\emph {\bibinfo {title} {Gauge Fields and Composite Fermions in
  Bilayer Quantum Hall Systems}},\ \href@noop {} {Ph.D. thesis},\ \bibinfo
  {school} {Florida State University} (\bibinfo {year} {2014})\BibitemShut
  {NoStop}%
\bibitem [{\citenamefont {Cipri}\ and\ \citenamefont
  {Bonesteel}(2014)}]{CipriBonesteel}%
  \BibitemOpen
  \bibfield  {author} {\bibinfo {author} {\bibfnamefont {R.}~\bibnamefont
  {Cipri}}\ and\ \bibinfo {author} {\bibfnamefont {N.~E.}\ \bibnamefont
  {Bonesteel}},\ }\href {\doibase 10.1103/PhysRevB.89.085109} {\bibfield
  {journal} {\bibinfo  {journal} {Phys. Rev. B}\ }\textbf {\bibinfo {volume}
  {89}},\ \bibinfo {pages} {085109} (\bibinfo {year} {2014})}\BibitemShut
  {NoStop}%
\bibitem [{\citenamefont {Papic}(2010)}]{papicThesis}%
  \BibitemOpen
  \bibfield  {author} {\bibinfo {author} {\bibfnamefont {Z.}~\bibnamefont
  {Papic}},\ }\emph {\bibinfo {title} {Fractional quantum Hall effect in
  multicomponent systems}},\ \href@noop {} {Ph.D. thesis},\ \bibinfo  {school}
  {Universit{\'e} Paris Sud-Paris XI} (\bibinfo {year} {2010})\BibitemShut
  {NoStop}%
\bibitem [{\citenamefont {Doretto}\ \emph {et~al.}(2012)\citenamefont
  {Doretto}, \citenamefont {Morais~Smith},\ and\ \citenamefont
  {Caldeira}}]{Bosonization1}%
  \BibitemOpen
  \bibfield  {author} {\bibinfo {author} {\bibfnamefont {R.~L.}\ \bibnamefont
  {Doretto}}, \bibinfo {author} {\bibfnamefont {C.}~\bibnamefont
  {Morais~Smith}}, \ and\ \bibinfo {author} {\bibfnamefont {A.~O.}\
  \bibnamefont {Caldeira}},\ }\href {\doibase 10.1103/PhysRevB.86.035326}
  {\bibfield  {journal} {\bibinfo  {journal} {Phys. Rev. B}\ }\textbf {\bibinfo
  {volume} {86}},\ \bibinfo {pages} {035326} (\bibinfo {year}
  {2012})}\BibitemShut {NoStop}%
\bibitem [{\citenamefont {Doretto}\ \emph {et~al.}(2006)\citenamefont
  {Doretto}, \citenamefont {Caldeira},\ and\ \citenamefont
  {Smith}}]{Bosonization4}%
  \BibitemOpen
  \bibfield  {author} {\bibinfo {author} {\bibfnamefont {R.~L.}\ \bibnamefont
  {Doretto}}, \bibinfo {author} {\bibfnamefont {A.~O.}\ \bibnamefont
  {Caldeira}}, \ and\ \bibinfo {author} {\bibfnamefont {C.~M.}\ \bibnamefont
  {Smith}},\ }\href {\doibase 10.1103/PhysRevLett.97.186401} {\bibfield
  {journal} {\bibinfo  {journal} {Phys. Rev. Lett.}\ }\textbf {\bibinfo
  {volume} {97}},\ \bibinfo {pages} {186401} (\bibinfo {year}
  {2006})}\BibitemShut {NoStop}%
\bibitem [{\citenamefont {Halperin}(2020)}]{Halperin2020}%
  \BibitemOpen
  \bibfield  {author} {\bibinfo {author} {\bibfnamefont {B.~I.}\ \bibnamefont
  {Halperin}},\ }in\ \href@noop {} {\emph {\bibinfo {booktitle} {Fractional
  Quantum Hall Effects: New Developments}}},\ \bibinfo {editor} {edited by\
  \bibinfo {editor} {\bibfnamefont {B.~I.}\ \bibnamefont {Halperin}}\ and\
  \bibinfo {editor} {\bibfnamefont {J.~K.}\ \bibnamefont {Jain}}}\ (\bibinfo
  {publisher} {World Scientific},\ \bibinfo {year} {2020})\ pp.\ \bibinfo
  {pages} {79--132}\BibitemShut {NoStop}%
\bibitem [{\citenamefont {{Liu}}\ \emph {et~al.}(2020)\citenamefont {{Liu}},
  \citenamefont {{Li}}, \citenamefont {{Watanabe}}, \citenamefont
  {{Taniguchi}}, \citenamefont {{Hone}}, \citenamefont {{Halperin}},
  \citenamefont {{Kim}},\ and\ \citenamefont {{Dean}}}]{Crossover}%
  \BibitemOpen
  \bibfield  {author} {\bibinfo {author} {\bibfnamefont {X.}~\bibnamefont
  {{Liu}}}, \bibinfo {author} {\bibfnamefont {J.~I.~A.}\ \bibnamefont {{Li}}},
  \bibinfo {author} {\bibfnamefont {K.}~\bibnamefont {{Watanabe}}}, \bibinfo
  {author} {\bibfnamefont {T.}~\bibnamefont {{Taniguchi}}}, \bibinfo {author}
  {\bibfnamefont {J.}~\bibnamefont {{Hone}}}, \bibinfo {author} {\bibfnamefont
  {B.~I.}\ \bibnamefont {{Halperin}}}, \bibinfo {author} {\bibfnamefont
  {P.}~\bibnamefont {{Kim}}}, \ and\ \bibinfo {author} {\bibfnamefont {C.~R.}\
  \bibnamefont {{Dean}}},\ }\href@noop {} {\bibfield  {journal} {\bibinfo
  {journal} {arXiv e-prints}\ ,\ \bibinfo {eid} {see in particular
  Supplementary Material}} (\bibinfo {year} {2020})},\ \Eprint
  {http://arxiv.org/abs/2012.05916} {arXiv:2012.05916 [cond-mat.mes-hall]}
  \BibitemShut {NoStop}%
\bibitem [{\citenamefont {Son}(2015)}]{Son}%
  \BibitemOpen
  \bibfield  {author} {\bibinfo {author} {\bibfnamefont {D.~T.}\ \bibnamefont
  {Son}},\ }\href {\doibase 10.1103/PhysRevX.5.031027} {\bibfield  {journal}
  {\bibinfo  {journal} {Phys. Rev. X}\ }\textbf {\bibinfo {volume} {5}},\
  \bibinfo {pages} {031027} (\bibinfo {year} {2015})}\BibitemShut {NoStop}%
\bibitem [{\citenamefont {Nguyen}\ \emph {et~al.}(2018)\citenamefont {Nguyen},
  \citenamefont {Golkar}, \citenamefont {Roberts},\ and\ \citenamefont
  {Son}}]{Dirac-Full}%
  \BibitemOpen
  \bibfield  {author} {\bibinfo {author} {\bibfnamefont {D.~X.}\ \bibnamefont
  {Nguyen}}, \bibinfo {author} {\bibfnamefont {S.}~\bibnamefont {Golkar}},
  \bibinfo {author} {\bibfnamefont {M.~M.}\ \bibnamefont {Roberts}}, \ and\
  \bibinfo {author} {\bibfnamefont {D.~T.}\ \bibnamefont {Son}},\ }\href
  {\doibase 10.1103/PhysRevB.97.195314} {\bibfield  {journal} {\bibinfo
  {journal} {Phys. Rev. B}\ }\textbf {\bibinfo {volume} {97}},\ \bibinfo
  {pages} {195314} (\bibinfo {year} {2018})}\BibitemShut {NoStop}%
\bibitem [{\citenamefont {Geraedts}\ \emph {et~al.}(2016)\citenamefont
  {Geraedts}, \citenamefont {Zaletel}, \citenamefont {Mong}, \citenamefont
  {Metlitski}, \citenamefont {Vishwanath},\ and\ \citenamefont
  {Motrunich}}]{DMRG_half_filled}%
  \BibitemOpen
  \bibfield  {author} {\bibinfo {author} {\bibfnamefont {S.~D.}\ \bibnamefont
  {Geraedts}}, \bibinfo {author} {\bibfnamefont {M.~P.}\ \bibnamefont
  {Zaletel}}, \bibinfo {author} {\bibfnamefont {R.~S.~K.}\ \bibnamefont
  {Mong}}, \bibinfo {author} {\bibfnamefont {M.~A.}\ \bibnamefont {Metlitski}},
  \bibinfo {author} {\bibfnamefont {A.}~\bibnamefont {Vishwanath}}, \ and\
  \bibinfo {author} {\bibfnamefont {O.~I.}\ \bibnamefont {Motrunich}},\ }\href
  {\doibase 10.1126/science.aad4302} {\bibfield  {journal} {\bibinfo  {journal}
  {Science}\ }\textbf {\bibinfo {volume} {352}},\ \bibinfo {pages} {197}
  (\bibinfo {year} {2016})}\BibitemShut {NoStop}%
\bibitem [{\citenamefont {Rezayi}\ and\ \citenamefont
  {Haldane}(2000)}]{RezayiHaldane}%
  \BibitemOpen
  \bibfield  {author} {\bibinfo {author} {\bibfnamefont {E.~H.}\ \bibnamefont
  {Rezayi}}\ and\ \bibinfo {author} {\bibfnamefont {F.~D.~M.}\ \bibnamefont
  {Haldane}},\ }\href {\doibase 10.1103/PhysRevLett.84.4685} {\bibfield
  {journal} {\bibinfo  {journal} {Phys. Rev. Lett.}\ }\textbf {\bibinfo
  {volume} {84}},\ \bibinfo {pages} {4685} (\bibinfo {year}
  {2000})}\BibitemShut {NoStop}%
\bibitem [{\citenamefont {Parish}()}]{BCS_BEC}%
  \BibitemOpen
  \bibfield  {author} {\bibinfo {author} {\bibfnamefont {M.~M.}\ \bibnamefont
  {Parish}},\ }\enquote {\bibinfo {title} {The {BCS-BEC} {Crossover}},}\ in\
  \href {https://www.worldscientific.com/doi/abs/10.1142/9781783264766_0009}
  {\emph {\bibinfo {booktitle} {Quantum Gas Experiments}}},\ Chap.\ \bibinfo
  {chapter} {Chapter 9}, pp.\ \bibinfo {pages} {179--197}\BibitemShut {NoStop}%
\bibitem [{\citenamefont {Chen}\ \emph {et~al.}(2005)\citenamefont {Chen},
  \citenamefont {Stajic}, \citenamefont {Tan},\ and\ \citenamefont
  {Levin}}]{CHEN20051}%
  \BibitemOpen
  \bibfield  {author} {\bibinfo {author} {\bibfnamefont {Q.}~\bibnamefont
  {Chen}}, \bibinfo {author} {\bibfnamefont {J.}~\bibnamefont {Stajic}},
  \bibinfo {author} {\bibfnamefont {S.}~\bibnamefont {Tan}}, \ and\ \bibinfo
  {author} {\bibfnamefont {K.}~\bibnamefont {Levin}},\ }\href {\doibase
  https://doi.org/10.1016/j.physrep.2005.02.005} {\bibfield  {journal}
  {\bibinfo  {journal} {Physics Reports}\ }\textbf {\bibinfo {volume} {412}},\
  \bibinfo {pages} {1} (\bibinfo {year} {2005})}\BibitemShut {NoStop}%
\bibitem [{\citenamefont {Yang}(2001)}]{KunYang}%
  \BibitemOpen
  \bibfield  {author} {\bibinfo {author} {\bibfnamefont {K.}~\bibnamefont
  {Yang}},\ }\href {\doibase 10.1103/PhysRevLett.87.056802} {\bibfield
  {journal} {\bibinfo  {journal} {Phys. Rev. Lett.}\ }\textbf {\bibinfo
  {volume} {87}},\ \bibinfo {pages} {056802} (\bibinfo {year}
  {2001})}\BibitemShut {NoStop}%
\bibitem [{\citenamefont {Nguyen}\ \emph {et~al.}(2017)\citenamefont {Nguyen},
  \citenamefont {Can},\ and\ \citenamefont {Gromov}}]{PH}%
  \BibitemOpen
  \bibfield  {author} {\bibinfo {author} {\bibfnamefont {D.~X.}\ \bibnamefont
  {Nguyen}}, \bibinfo {author} {\bibfnamefont {T.}~\bibnamefont {Can}}, \ and\
  \bibinfo {author} {\bibfnamefont {A.}~\bibnamefont {Gromov}},\ }\href
  {\doibase 10.1103/PhysRevLett.118.206602} {\bibfield  {journal} {\bibinfo
  {journal} {Phys. Rev. Lett.}\ }\textbf {\bibinfo {volume} {118}},\ \bibinfo
  {pages} {206602} (\bibinfo {year} {2017})}\BibitemShut {NoStop}%
\bibitem [{\citenamefont {Wagner}\ \emph {et~al.}()\citenamefont {Wagner},
  \citenamefont {Nguyen}, \citenamefont {Simon},\ and\ \citenamefont
  {Halperin}}]{Supplement}%
  \BibitemOpen
  \bibfield  {author} {\bibinfo {author} {\bibfnamefont {G.}~\bibnamefont
  {Wagner}}, \bibinfo {author} {\bibfnamefont {D.~X.}\ \bibnamefont {Nguyen}},
  \bibinfo {author} {\bibfnamefont {S.~H.}\ \bibnamefont {Simon}}, \ and\
  \bibinfo {author} {\bibfnamefont {B.~I.}\ \bibnamefont {Halperin}},\
  }\href@noop {} {\bibinfo  {journal} {Supplementary Material to this
  article.}\ }\BibitemShut {NoStop}%
\bibitem [{\citenamefont {Jain}\ and\ \citenamefont
  {Kamilla}(1997)}]{Kamilla_Jain}%
  \BibitemOpen
\bibfield  {journal} {  }\bibfield  {author} {\bibinfo {author} {\bibfnamefont
  {J.~K.}\ \bibnamefont {Jain}}\ and\ \bibinfo {author} {\bibfnamefont {R.~K.}\
  \bibnamefont {Kamilla}},\ }\href {\doibase 10.1103/PhysRevB.55.R4895}
  {\bibfield  {journal} {\bibinfo  {journal} {Phys. Rev. B}\ }\textbf {\bibinfo
  {volume} {55}},\ \bibinfo {pages} {R4895} (\bibinfo {year}
  {1997})}\BibitemShut {NoStop}%
\bibitem [{\citenamefont {Kamilla}(1997)}]{Kamilla_thesis}%
  \BibitemOpen
  \bibfield  {author} {\bibinfo {author} {\bibfnamefont {R.~K.}\ \bibnamefont
  {Kamilla}},\ }\emph {\bibinfo {title} {Composite Fermions: Physics of
  2-Dimensional Electron Systems Under Strong Magnetic Fields}},\ \href@noop {}
  {Ph.D. thesis},\ \bibinfo  {school} {Stony Brook University} (\bibinfo {year}
  {1997})\BibitemShut {NoStop}%
\bibitem [{\citenamefont {Xiang}\ \emph {et~al.}(1997)\citenamefont {Xiang},
  \citenamefont {Sun}, \citenamefont {Fan},\ and\ \citenamefont {Gong}}]{DA}%
  \BibitemOpen
  \bibfield  {author} {\bibinfo {author} {\bibfnamefont {Y.}~\bibnamefont
  {Xiang}}, \bibinfo {author} {\bibfnamefont {D.}~\bibnamefont {Sun}}, \bibinfo
  {author} {\bibfnamefont {W.}~\bibnamefont {Fan}}, \ and\ \bibinfo {author}
  {\bibfnamefont {X.}~\bibnamefont {Gong}},\ }\href {\doibase
  https://doi.org/10.1016/S0375-9601(97)00474-X} {\bibfield  {journal}
  {\bibinfo  {journal} {Physics Letters A}\ }\textbf {\bibinfo {volume}
  {233}},\ \bibinfo {pages} {216} (\bibinfo {year} {1997})}\BibitemShut
  {NoStop}%
\bibitem [{\citenamefont {Rezayi}\ and\ \citenamefont {Read}(1994)}]{Hund}%
  \BibitemOpen
  \bibfield  {author} {\bibinfo {author} {\bibfnamefont {E.}~\bibnamefont
  {Rezayi}}\ and\ \bibinfo {author} {\bibfnamefont {N.}~\bibnamefont {Read}},\
  }\href {\doibase 10.1103/PhysRevLett.72.900} {\bibfield  {journal} {\bibinfo
  {journal} {Phys. Rev. Lett.}\ }\textbf {\bibinfo {volume} {72}},\ \bibinfo
  {pages} {900} (\bibinfo {year} {1994})}\BibitemShut {NoStop}%
\bibitem [{\citenamefont {Kamburov}\ \emph {et~al.}(2014)\citenamefont
  {Kamburov}, \citenamefont {Liu}, \citenamefont {Mueed}, \citenamefont
  {Shayegan}, \citenamefont {Pfeiffer}, \citenamefont {West},\ and\
  \citenamefont {Baldwin}}]{minority_carrier}%
  \BibitemOpen
  \bibfield  {author} {\bibinfo {author} {\bibfnamefont {D.}~\bibnamefont
  {Kamburov}}, \bibinfo {author} {\bibfnamefont {Y.}~\bibnamefont {Liu}},
  \bibinfo {author} {\bibfnamefont {M.~A.}\ \bibnamefont {Mueed}}, \bibinfo
  {author} {\bibfnamefont {M.}~\bibnamefont {Shayegan}}, \bibinfo {author}
  {\bibfnamefont {L.~N.}\ \bibnamefont {Pfeiffer}}, \bibinfo {author}
  {\bibfnamefont {K.~W.}\ \bibnamefont {West}}, \ and\ \bibinfo {author}
  {\bibfnamefont {K.~W.}\ \bibnamefont {Baldwin}},\ }\href {\doibase
  10.1103/PhysRevLett.113.196801} {\bibfield  {journal} {\bibinfo  {journal}
  {Phys. Rev. Lett.}\ }\textbf {\bibinfo {volume} {113}},\ \bibinfo {pages}
  {196801} (\bibinfo {year} {2014})}\BibitemShut {NoStop}%
\bibitem [{\citenamefont {Hossain}\ \emph {et~al.}(2020)\citenamefont
  {Hossain}, \citenamefont {Mueed}, \citenamefont {Ma}, \citenamefont
  {Villegas~Rosales}, \citenamefont {Chung}, \citenamefont {Pfeiffer},
  \citenamefont {West}, \citenamefont {Baldwin},\ and\ \citenamefont
  {Shayegan}}]{minority_carrier2}%
  \BibitemOpen
  \bibfield  {author} {\bibinfo {author} {\bibfnamefont {M.~S.}\ \bibnamefont
  {Hossain}}, \bibinfo {author} {\bibfnamefont {M.~A.}\ \bibnamefont {Mueed}},
  \bibinfo {author} {\bibfnamefont {M.~K.}\ \bibnamefont {Ma}}, \bibinfo
  {author} {\bibfnamefont {K.~A.}\ \bibnamefont {Villegas~Rosales}}, \bibinfo
  {author} {\bibfnamefont {Y.~J.}\ \bibnamefont {Chung}}, \bibinfo {author}
  {\bibfnamefont {L.~N.}\ \bibnamefont {Pfeiffer}}, \bibinfo {author}
  {\bibfnamefont {K.~W.}\ \bibnamefont {West}}, \bibinfo {author}
  {\bibfnamefont {K.~W.}\ \bibnamefont {Baldwin}}, \ and\ \bibinfo {author}
  {\bibfnamefont {M.}~\bibnamefont {Shayegan}},\ }\href {\doibase
  10.1103/PhysRevLett.125.046601} {\bibfield  {journal} {\bibinfo  {journal}
  {Phys. Rev. Lett.}\ }\textbf {\bibinfo {volume} {125}},\ \bibinfo {pages}
  {046601} (\bibinfo {year} {2020})}\BibitemShut {NoStop}%
\bibitem [{\citenamefont {Champagne}\ \emph {et~al.}(2008)\citenamefont
  {Champagne}, \citenamefont {Finck}, \citenamefont {Eisenstein}, \citenamefont
  {Pfeiffer},\ and\ \citenamefont {West}}]{Experiments2}%
  \BibitemOpen
  \bibfield  {author} {\bibinfo {author} {\bibfnamefont {A.~R.}\ \bibnamefont
  {Champagne}}, \bibinfo {author} {\bibfnamefont {A.~D.~K.}\ \bibnamefont
  {Finck}}, \bibinfo {author} {\bibfnamefont {J.~P.}\ \bibnamefont
  {Eisenstein}}, \bibinfo {author} {\bibfnamefont {L.~N.}\ \bibnamefont
  {Pfeiffer}}, \ and\ \bibinfo {author} {\bibfnamefont {K.~W.}\ \bibnamefont
  {West}},\ }\href {\doibase 10.1103/PhysRevB.78.205310} {\bibfield  {journal}
  {\bibinfo  {journal} {Phys. Rev. B}\ }\textbf {\bibinfo {volume} {78}},\
  \bibinfo {pages} {205310} (\bibinfo {year} {2008})}\BibitemShut {NoStop}%
\bibitem [{\citenamefont {Clarke}\ \emph {et~al.}(2004)\citenamefont {Clarke},
  \citenamefont {Micolich}, \citenamefont {Hamilton}, \citenamefont {Simmons},
  \citenamefont {Pepper},\ and\ \citenamefont {Ritchie}}]{Experiments3}%
  \BibitemOpen
  \bibfield  {author} {\bibinfo {author} {\bibfnamefont {W.}~\bibnamefont
  {Clarke}}, \bibinfo {author} {\bibfnamefont {A.}~\bibnamefont {Micolich}},
  \bibinfo {author} {\bibfnamefont {A.}~\bibnamefont {Hamilton}}, \bibinfo
  {author} {\bibfnamefont {M.}~\bibnamefont {Simmons}}, \bibinfo {author}
  {\bibfnamefont {M.}~\bibnamefont {Pepper}}, \ and\ \bibinfo {author}
  {\bibfnamefont {D.}~\bibnamefont {Ritchie}},\ }\href {\doibase
  https://doi.org/10.1016/j.physe.2003.11.211} {\bibfield  {journal} {\bibinfo
  {journal} {Physica E: Low-dimensional Systems and Nanostructures}\ }\textbf
  {\bibinfo {volume} {22}},\ \bibinfo {pages} {40 } (\bibinfo {year}
  {2004})}\BibitemShut {NoStop}%
\bibitem [{\citenamefont {Spielman}\ \emph {et~al.}(2004)\citenamefont
  {Spielman}, \citenamefont {Kellogg}, \citenamefont {Eisenstein},
  \citenamefont {Pfeiffer},\ and\ \citenamefont {West}}]{Experiments4}%
  \BibitemOpen
  \bibfield  {author} {\bibinfo {author} {\bibfnamefont {I.~B.}\ \bibnamefont
  {Spielman}}, \bibinfo {author} {\bibfnamefont {M.}~\bibnamefont {Kellogg}},
  \bibinfo {author} {\bibfnamefont {J.~P.}\ \bibnamefont {Eisenstein}},
  \bibinfo {author} {\bibfnamefont {L.~N.}\ \bibnamefont {Pfeiffer}}, \ and\
  \bibinfo {author} {\bibfnamefont {K.~W.}\ \bibnamefont {West}},\ }\href
  {\doibase 10.1103/PhysRevB.70.081303} {\bibfield  {journal} {\bibinfo
  {journal} {Phys. Rev. B}\ }\textbf {\bibinfo {volume} {70}},\ \bibinfo
  {pages} {081303} (\bibinfo {year} {2004})}\BibitemShut {NoStop}%
\bibitem [{Note1()}]{Note1}%
  \BibitemOpen
  \bibinfo {note} {Note that the phases of the Jastrow factors
  $(z_i-z_j)^2(w_i-w_j)^{*2}$ cancel in the tightly bound limit where the CFs
  and anti-CFs are at the same position.}\BibitemShut {Stop}%
\bibitem [{Note2()}]{Note2}%
  \BibitemOpen
  \bibinfo {note} {Note that the phase factor $e^{i q \varphi _{j}}$ is missing
  from Eq. (2.9) in\cite {Kamilla_thesis}.}\BibitemShut {Stop}%
\bibitem [{\citenamefont {M\"oller}\ and\ \citenamefont
  {Simon}(2005)}]{negative_q}%
  \BibitemOpen
  \bibfield  {author} {\bibinfo {author} {\bibfnamefont {G.}~\bibnamefont
  {M\"oller}}\ and\ \bibinfo {author} {\bibfnamefont {S.~H.}\ \bibnamefont
  {Simon}},\ }\href {\doibase 10.1103/PhysRevB.72.045344} {\bibfield  {journal}
  {\bibinfo  {journal} {Phys. Rev. B}\ }\textbf {\bibinfo {volume} {72}},\
  \bibinfo {pages} {045344} (\bibinfo {year} {2005})}\BibitemShut {NoStop}%
\bibitem [{\citenamefont {Wu}\ and\ \citenamefont {Yang}(1977)}]{Identities}%
  \BibitemOpen
  \bibfield  {author} {\bibinfo {author} {\bibfnamefont {T.~T.}\ \bibnamefont
  {Wu}}\ and\ \bibinfo {author} {\bibfnamefont {C.~N.}\ \bibnamefont {Yang}},\
  }\href {\doibase 10.1103/PhysRevD.16.1018} {\bibfield  {journal} {\bibinfo
  {journal} {Phys. Rev. D}\ }\textbf {\bibinfo {volume} {16}},\ \bibinfo
  {pages} {1018} (\bibinfo {year} {1977})}\BibitemShut {NoStop}%
\bibitem [{\citenamefont {Liu}\ \emph {et~al.}(2021)\citenamefont {Liu},
  \citenamefont {Balram}, \citenamefont {Papi\ifmmode~\acute{c}\else
  \'{c}\fi{}},\ and\ \citenamefont {Gromov}}]{Balram}%
  \BibitemOpen
  \bibfield  {author} {\bibinfo {author} {\bibfnamefont {Z.}~\bibnamefont
  {Liu}}, \bibinfo {author} {\bibfnamefont {A.~C.}\ \bibnamefont {Balram}},
  \bibinfo {author} {\bibfnamefont {Z.}~\bibnamefont
  {Papi\ifmmode~\acute{c}\else \'{c}\fi{}}}, \ and\ \bibinfo {author}
  {\bibfnamefont {A.}~\bibnamefont {Gromov}},\ }\href {\doibase
  10.1103/PhysRevLett.126.076604} {\bibfield  {journal} {\bibinfo  {journal}
  {Phys. Rev. Lett.}\ }\textbf {\bibinfo {volume} {126}},\ \bibinfo {pages}
  {076604} (\bibinfo {year} {2021})}\BibitemShut {NoStop}%
\end{thebibliography}%

\newpage
\pagebreak
\clearpage
\appendix
\begin{widetext}
	\begin{center}
		\textbf{\large --- Supplementary Material ---\\ $s$-wave paired composite-fermion electron-hole trial state for quantum Hall bilayers with $\nu=1$}\\
		\medskip
		\text{Glenn Wagner, Dung X. Nguyen, Steven H. Simon and Bertrand I. Halperin }
	\end{center}
	
		\setcounter{equation}{0}
	\setcounter{figure}{0}
	\setcounter{table}{0}
	\setcounter{page}{1}
	\makeatletter
	\renewcommand{\theequation}{S\arabic{equation}}
	\renewcommand{\thefigure}{S\arabic{figure}}
	\renewcommand{\bibnumfmt}[1]{[S#1]}

\section{\uppercase{Composite fermion wavefunctions}}

 We present here a detailed discussionof how we have constructed trial wavefunctions in the spherical geometry, for the various paired states we consider.

The single particle electron eigenstates on the sphere are the monopole harmonics $Y_{q, n, m}(\Omega)$, where the total flux through the sphere is $N_\phi=2q$ ($q$ can be integer or half-integer), $n=0,1,2,\dots$ is the LL index (0 for LLL) and $m=-q-n,-q-n+1, \ldots, q+n$ labels the $2(q+n)+1$ degenerate states within a LL. We can define the angular momentum quantum number $l=q+n$, such that $l=q,q+1,\dots$ and $m=-l,\dots,l$. From\cite{Kamilla_Jain} the wavefunction for electron $j$ is
\begin{equation}\label{eq:monopole_harmonics}
\begin{aligned}
Y_{q, n, m}\left(\Omega_{j}\right)=& N_{q n m}(-1)^{q+n-m} e^{i q \varphi_{j}} u_{j}^{q+m} v_{j}^{q-m}\\
& \times \sum_{s=0}^{n}(-1)^{s} \left(\begin{array}{c}
n \\
s
\end{array}\right)\left(\begin{array}{c}
2 q+n \\
q+n-m-s
\end{array}\right)\left(v_{j}^{*} v_{j}\right)^{n-s}\left(u_{j}^{*} u_{j}\right)^{s},
\end{aligned}
\end{equation}
where the spinor coordinates are
\begin{equation}
u_j=\cos (\theta_j / 2) e^{-i \varphi_j / 2} \text { and } v_j=\sin (\theta_j / 2) e^{i \varphi_j / 2}
\end{equation}
in terms of the usual polar coordinates $\Omega_j=(\theta_j,\varphi_j)$ on the sphere. The normalization constant $N_{q n m}$ is 
\begin{equation}N_{qnm}=\sqrt{\frac{2q+2 n+1}{4 \pi} \frac{(q+n-m) !(q+n+m) !}{n !(2q+n) !}}=\sqrt{\frac{2l+1}{4\pi}\frac{(l-m)!(l+m)!}{n!(l+q)!}}.
\end{equation}
Let us assume we have $N_1$ electrons in a given layer. We perform the flux attachment procedure for CFs by multiplying the many-body wavefunction by the Jastrow factor
\begin{equation}
\mathcal{J}=\prod_{j} J_{j}=\prod_{j < k}\left(u_{j} v_{k}-v_{j} u_{k}\right)^{2} e^{i \left(\varphi_{j}+\varphi_{k}\right)}\equiv\prod_{j < k}(\Omega_j-\Omega_k)^2,
\end{equation}
where the have defined the short-hand notation $\prod_{j < k}(\Omega_j-\Omega_k)^2$ for the Jastrow factors. The CFs now experience flux $N_\phi^\textrm{eff}=2Q$, where $Q=q-(N_1-1)$. To write down the state appropriate for a single layer, we fill the CF orbitals and if we have a partially filled shell, we fill it according to Hund's rule\cite{Hund} (maximize total angular momentum). Once we have decided which states to fill, we can write down a single Slater determinant and attach the flux quanta via the Jastrow factor
\begin{equation}
\Psi_\textrm{CFL}(\{\Omega\})=\mathcal{P}_\textrm{LLL}\bigg[\mathcal{J} \operatorname{det}\left[Y_{i}\left(\Omega_{j}\right)\right]\bigg],
\end{equation}
where $i$ is a short hand for the indices $(Q,n,m)$ for the filled states. We use the trick from the paper by Jain and Kamilla\cite{Kamilla_Jain} to project the single particle states (together with the Jastrow factor $J_j$) to the LLL. The single Slater determinant of states that are all in the LLL will be in the LLL as well. The LLL projection is otherwise computationally intractable. The resulting states have a high overlap with the states obtained when the LLL projection is done properly. The projected single particle states are defined via
\begin{equation}
\tilde Y_{i}\left(\Omega_{j}\right) J_{j}=\mathcal{P}_\textrm{LLL}[ Y_{i}\left(\Omega_{j}\right) J_{j}]
\end{equation}
and the explicit expression is\cite{Kamilla_Jain,Kamilla_thesis} \footnote{Note that the phase factor $e^{i q \varphi_{j}}$ is missing from Eq. (2.9) in\cite{Kamilla_thesis}.}
\begin{equation}
\begin{aligned}
\tilde{Y}_{Q, n, m}\left(\Omega_{j}\right)=& N_{Q n m}(-1)^{Q+n-m} \frac{(2 q+1) !}{(2 q+n+1) !} u_{j}^{Q+m} v_{j}^{Q-m} e^{i Q \varphi_{j}} \\
& \times\sum_{s=0}^{n}(-1)^{s}\left(\begin{array}{c}
n \\
s
\end{array}\right) \left(\begin{array}{c}
2 Q+n \\
Q+n-m-s
\end{array}\right) v_{j}^{n-s} u_{j}^{s} R_{j}^{s, n-s},
\end{aligned}
\end{equation}
where   
\begin{equation}R_{j}^{s, n-s}=\mathbf{U}_{j}^{s} \mathbf{V}_{j}^{n-s} 1\end{equation}
with
\begin{equation}\begin{aligned}
&\mathbf{U}_{j}=\sum_{k\neq j}\frac{v_{k}}{u_{j} v_{k}-v_{j} u_{k}}+\frac{\partial}{\partial u_{j}}\\
&\mathbf{V}_{j}=\sum_{k\neq j} \frac{-u_{k}}{u_{j} v_{k}-v_{j} u_{k}}+\frac{\partial}{\partial v_{j}}.
\end{aligned}\end{equation}
A good approximation for the CFL is then
\begin{equation}
\tilde \Psi_\textrm{CFL}(\{\Omega\})=\prod_{k < l}(\Omega_k-\Omega_l)^2 \det\left[\tilde{Y}_{i}\left(\Omega_{j}\right)\right].
\end{equation}
When we have two layers, we write down the filled shell/Hund's rule state for the two individual layers and then we pair the two layers with the appropriate Clebsch-Gordan coefficients to form an $L^2=0$ state.

\section{\uppercase{Paired states}}

\subsection{$p$-wave BCS state}
Let us first recapitulate the $p$-wave paired BCS state. We work in the sector with total particle number $N=2q+1$, where $q$ is half-integer, such that $N$ is even and the number of electrons in each layer is $N_1=N_\uparrow=N_\downarrow=\frac{N}{2}$. Electron coordinates in the top layer will be $\Omega_i^\uparrow$ with $i=1,\dots,\frac{N}{2}$ and electron coordinates in the bottom layer will be $\Omega_i^\downarrow$ with $i=1,\dots,\frac{N}{2}$. We attach flux to the electrons in the top layer and bottom layer. The net flux experienced by CFs in both layers is $Q_{\uparrow,\downarrow}=q-(N_{\uparrow,\downarrow}-1)=\frac{1}{2}$. For the $p$-wave pairing of CFs as in\cite{Simon1,Simon2} the pairing wavefunction is
\begin{equation}
    G(\Omega_{i}^{\uparrow},\Omega_{j}^{\downarrow})=\sum_{n=0}^{N_\textrm{LL}-1}g_{n} \sum_{m=-(n+1/2)}^{n+1/2}(-1)^{\frac{1}{2}+m}  {Y}_{\frac{1}{2}, n, m}\left(\Omega_{i}^{\uparrow}\right) {Y}_{\frac{1}{2}, n,-m}\left(\Omega_{j}^{\downarrow}\right).
\end{equation}
$N_\textrm{LL}$ is the number of CF Landau levels (LLs) that are included. The LLL wavefunction is then
\begin{equation}
\Psi_{\textrm{BCS},p}(\{\Omega^\uparrow\},\{\Omega^\downarrow\})=\mathcal{P}_\textrm{LLL} \prod_{i < j}[(\Omega^\uparrow_i-\Omega^\uparrow_j)^2 (\Omega^\downarrow_i-\Omega^\downarrow_j)^2] \mathrm{det} [G(\Omega_{i}^{\uparrow},\Omega_{j}^{\downarrow})].
\end{equation}
However, computing this wavefunction will be too expensive numerically and hence we resort to the procedure described in\cite{Kamilla_Jain}. The wavefunction is then
\begin{equation}
\tilde \Psi_{\textrm{BCS},p}(\{\Omega^\uparrow\},\{\Omega^\downarrow\})=\prod_{i < j}[ (\Omega^\uparrow_i-\Omega^\uparrow_j)^2 (\Omega^\downarrow_i-\Omega^\downarrow_j)^2] \mathrm{det} [\tilde G(\Omega_{i}^{\uparrow},\Omega_{j}^{\downarrow})]
\end{equation}
with
\begin{equation}
\tilde G(\Omega_{i}^{\uparrow},\Omega_{j}^{\downarrow})=\bigg[ \sum_{n=0}^{N_\textrm{LL}-1} \sum_{m=-(n+1/2)}^{n+1/2}(-1)^{\frac{1}{2}+m} g_{n} \tilde{Y}_{\frac{1}{2}, n, m}\left(\Omega_{i}^{\uparrow}\right) \tilde{Y}_{\frac{1}{2}, n,-m}\left(\Omega_{j}^{\downarrow}\right)\bigg].
\end{equation}

\subsection{New $s$-wave paired BCS state}
We now perform a particle-hole transformation on the bottom layer, such that $\Omega_1^{\uparrow},\cdots,\Omega_{N_1}^{\uparrow}$ are coordinates of electrons in the top layer and $\varpi_1^{\downarrow},\cdots,\varpi_{N_1}^{\downarrow}$ are coordinates of holes in the bottom layer. We attach flux to the electrons in the top layer to obtain CFs and we attach flux to the holes in the bottom layer to obtain anti-CFs. In the balanced case, both the CFs and the anti-CFs experience flux $Q=\frac{1}{2}$. The pairing function for pairing the CFs in the top layer with the anti-CFs in the bottom layer in the $s$-wave channel is
\begin{equation}
    G(\Omega_{i}^{\uparrow},\varpi_{j}^{\downarrow})=\sum_{n=0}^{N_\textrm{LL}-1} g_{n} \sum_{m=-(n+1/2)}^{n+1/2}{Y}_{\frac{1}{2}, n, m}\left(\Omega_{i}^{\uparrow}\right) {Y}^*_{\frac{1}{2}, n,m}\left(\varpi_{j}^{\downarrow}\right).
\end{equation}
and the trial wavefunction is obtained by adding the Jastrow factors and performing the LLL projection
\begin{equation}
    \Psi_{\textrm{BCS},s}(\{\Omega^{\uparrow}\},\{\varpi^{\downarrow}\})=\mathcal{P}_\textrm{LLL} \prod_{i < j}[(\Omega_{i}^{\uparrow}-\Omega_{j}^{\uparrow})^2 (\varpi_{i}^{\downarrow}-\varpi_{j}^{\downarrow})^{*2}] \mathrm{det} [G(\Omega_{i}^{\uparrow},\varpi_{j}^{\downarrow})].
\end{equation}
This wavefunction describes pairing of CFs and anti-CFs. This time after the Jain and Kamilla procedure
\begin{equation}\label{eq:s_wave}
\tilde \Psi_{\textrm{BCS},s}(\{\Omega^\uparrow\},\{\varpi^\downarrow\})=\prod_{i < j}[ (\Omega^\uparrow_i-\Omega^\uparrow_j)^2 (\varpi^\downarrow_i-\varpi^\downarrow_j)^{*2}] \mathrm{det} [\tilde G(\Omega_{i}^{\uparrow},\varpi_{j}^{\downarrow})]
\end{equation}
with
\begin{equation}
\tilde G(\Omega_{i}^{\uparrow},\varpi_{j}^{\downarrow})=\bigg[ \sum_{n=0}^{N_\textrm{LL}-1} \sum_{m=-(n+1/2)}^{n+1/2} g_{n} \tilde{Y}_{\frac{1}{2},n, m}\left(\Omega_{i}^{\uparrow}\right) \tilde{Y}^*_{\frac{1}{2}, n,m}\left(\varpi_{j}^{\downarrow}\right)\bigg].
\end{equation}
In practice we will have to truncate the number of variational parameters. Note that the variational parameters $\{g_n\}$ can always be chosen real.

We can also consider the imbalanced case, where we have $N_1=N_{\uparrow}$ electrons in the top layer and $N_{\downarrow}$ electrons in the bottom layer. The total number of electrons still satisfies $N=N_{\uparrow}+N_{\downarrow}=N_\phi+1$. We define the pseudospin as $2S_z=N_{\uparrow}-N_{\downarrow}$. Again we perform a particle-hole transformation on the bottom layer, such that $\Omega_1^{\uparrow},\cdots,\Omega_{N_1}^{\uparrow}$ are coordinates of electrons in the top layer and $\varpi_1^{\downarrow},\cdots,\varpi_{N_1}^{\downarrow}$ are coordinates of holes in the bottom layer. After the flux attachment procedure, the CFs in the top layer and the anti-CFs in the bottom layer feel an effective flux $Q=\frac{1}{2}-S_z$. The trial state is then

\begin{equation}
\tilde \Psi_{\textrm{BCS},s}(\{\Omega^\uparrow\},\{\varpi^\downarrow\})=\prod_{i < j}[ (\Omega^\uparrow_i-\Omega^\uparrow_j)^2 (\varpi^\downarrow_i-\varpi^\downarrow_j)^{*2}] \mathrm{det} [\tilde G(\Omega_{i}^{\uparrow},\varpi_{j}^{\downarrow})]
\end{equation}
with
\begin{equation}
\tilde G(\Omega_{i}^{\uparrow},\Omega_{j}^{\downarrow})=\bigg[ \sum_{n=0}^{N_\textrm{LL}-1} \sum_{m=-(n+Q)}^{n+Q} g_{n} \tilde{ Y}_{Q,n, m}\left(\Omega_{i}^{\uparrow}\right) \tilde{Y}^*_{Q, n,m}\left(\varpi_{j}^{\downarrow}\right)\bigg].
\end{equation}
We note that depending on whether we particle-hole conjugate the majority or minority layer, we may have $Q<0$ in which case we need to use the expression for $\tilde{Y}$ from Ref.~\cite{negative_q}.

\subsection{Symmetry considerations}
\label{sec:symm}
We use the identities for the monopole harmonics from\cite{Identities}. In particular
\begin{equation}
Y_{q, n, m}^{*}=(-1)^{q+m}Y_{-q, n,-m}
\end{equation}
and
\begin{equation}\label{eq:YY_identity}
\begin{aligned}
\sum_{m}  Y_{q,n, m}\left(\theta^{\prime}, \varphi^{\prime}\right) Y^*_{q^{\prime},n, m}(\theta, \varphi) =\sqrt{\frac{2(q+n)+1}{4 \pi}} Y_{q,n,-q'}(\theta_{12}, 0)  e^{i\left(q \varphi^{\prime}-q^{\prime} \varphi\right)} e^{-i\left(q \gamma'+q^{\prime} \gamma-q^{\prime} \pi\right)},
\end{aligned}
\end{equation}
where $\theta_{12}$ is the chord distance between the two particles and the angle $\gamma$ is as defined in\cite{Simon1}. First let us consider the symmetry properties of the $p$-wave paired state. Using these two identities, one can show (after setting $q=-q'=Q$ and doing some relabelling) that
\begin{equation}
\begin{aligned}
\sum_{m}(-1)^{Q+m} Y_{Q,n, m}\left(\theta^{\prime}, \varphi^{\prime}\right) Y_{Q,n,-m}(\theta, \varphi) 
= \sqrt{\frac{2(Q+n)+1}{4 \pi}} Y_{Q,n, -Q}\left(\theta_{12}, 0\right) e^{i Q\left(\varphi+\varphi^{\prime}\right)} e^{-i Q\left(\gamma-\gamma^{\prime}+\pi\right)}.
\end{aligned}
\end{equation}
Now exchange the two particles. Under the exchange (equivalently a half-rotation)
\begin{equation}\label{eq:exch}
    \gamma\to \gamma-\pi\qquad \textrm{and}\qquad \gamma'\to \gamma'+\pi
\end{equation}
such that the pairing wavefunction picks up a phase of 
 $e^{2\pi iQ}=(-1)^{2Q}$ under the exchange.   
For the balanced bilayer, we have $Q=\frac{1}{2}$ and so the sign is consistent with what one would require for a pairing wavefunction with odd $\ell$. $Y_{Q,n, -Q}\left(\theta_{12}, 0\right)\sim \theta_{12}^{2Q}$ when $\theta_{12}\to 0$ and so when the pair comes together, the wavefunction indeed vanishes in the correct way for $\ell=2Q=1$.

Now let us consider the proposed $s$-wave paired state. To do that set $q'=q=Q$ in \eqref{eq:YY_identity} such that
\begin{equation}
\begin{aligned}
\sum_{m} Y_{Q,n, m}\left(\theta^{\prime}, \varphi^{\prime}\right) Y^*_{Q,n, m}(\theta, \varphi) =\sqrt{\frac{2(Q+n)+1}{4 \pi}} Y_{Q,n,Q}(\theta_{12}, 0)  e^{iQ\left(\varphi^{\prime}-\varphi\right)} e^{-iQ\left( \gamma'+\gamma-\pi\right)}.
\end{aligned}
\end{equation}
Under the exchange \eqref{eq:exch}, there is no additional phase picked up. In addition $Y_{Q,n, Q}\left(\theta_{12}, 0\right)\to \textrm{const.}$ as $\theta_{12}\to0$. So indeed this is consistent with $s$-wave pairing. 

\subsection{111 state}

In terms of electron operators in both layers, the 111 state on the sphere can be written as the determinant of an $N\times N$ matrix
\begin{align}
\label{eq:111_ee}
    \Psi_{111}(\{\Omega^\uparrow\},\{\Omega^\downarrow\})&=\mathrm{det} [Y_{q,0,m_i}(\Omega_{j}^{\uparrow})\dots  Y_{q,0,m_i}(\Omega_{j}^{\downarrow})]\\
    &=\prod_{i < j}[(\Omega^\uparrow_i-\Omega^\uparrow_j) (\Omega^\downarrow_i-\Omega^\downarrow_j)]\prod_{i,j}(\Omega^\uparrow_i-\Omega^\downarrow_j),
\end{align}
where $m_i=-q,\dots,q$. In terms of electron coordinates in the top layer and hole operators in the bottom layer, we find
\begin{equation}
\label{eq:111_eh}
\Psi_{111}(\{\Omega^\uparrow\},\{\varpi^\downarrow\})=\textrm{det}\bigg[\sum_{m=-q}^{q}Y_{q,0,m}(\Omega^\uparrow_i)Y_{q,0,m}^*(\varpi^\downarrow_j)\bigg].
\end{equation}

\section{$p$-wave pairing in the imbalanced case}

 It is more natural to consider $s$-wave pairing of CF anti-CFs for the case of imbalanced layers, since the CF and anti-CF Fermi seas in opposite layers will have the same size. However, one can also consider the extension of the $p$-wave paired state to the imbalanced layer case for certain fillings. In particular, if the states fall on the Jain sequence, viz.~$\nu_\uparrow=\frac{p}{2p+1}$ and $\nu_\downarrow=\frac{p+1}{2p+1}$, then we can view the top layer as CFs filling the first $p$ LLs and the bottom layer as CFs filling the first $p+1$ LLs in an opposite effective magnetic field. We can now pair up CFs in the $n$th CF LL of the bottom layer with CFs in the $(n+1)$th LL of the top layer. In this way, we pair up all the CFs in the bottom layer with CFs in the top layer, for $n=0$ up to $n=p$,  leaving unpaired CFs in the $n=0$ CF LL of the top layer. This absence of pairing should have little effect on the total energy, (for large $p$)  since these electrons are far from the Fermi energy.

\subsection{Paired CB state}
In this section we consider a further trial state. We attach one flux quantum to electrons in the top layer to form composite bosons (CBs) and attach one flux quantum to holes in the bottom layer to form anti-CBs and then we pair the CBs and anti-CBs in the $s$-wave channel. This is similar to the construction proposed in \cite{ShouCheng}, except that we write the state on the sphere and introduce a larger number of variational parameters. As before, we perform a particle-hole transformation on the bottom layer, such that $\Omega_1^{\uparrow},\cdots,\Omega_{N_1}^{\uparrow}$ are coordinates of electrons in the top layer and $\varpi_1^{\downarrow},\cdots,\varpi_{N_1}^{\downarrow}$ are coordinates of holes in the bottom layer. After flux attachment, the CBs will experience flux $\mathcal Q=N_1$ in the balanced case. The pairing function for pairing CBs in the top layer and anti-CBs in the bottom layer in the $s$-wave channel is
\begin{equation}
    G(\Omega_{i}^{\uparrow},\varpi_{j}^{\downarrow})=\sum_{n=0}^{N_\textrm{LL}-1} g_{n} \sum_{m=-(n+\mathcal Q)}^{n+\mathcal Q}{Y}_{\mathcal Q, n, m}\left(\Omega_{i}^{\uparrow}\right) {Y}^*_{\mathcal Q, n,m}\left(\varpi_{j}^{\downarrow}\right).
\end{equation}
and including the Jastrow factors and performing the LLL projection we find the trial wavefunction
\begin{equation}
    \Psi_{\textrm{CB}}(\{\Omega^{\uparrow}\},\{\varpi^{\downarrow}\})=\mathcal{P}_\textrm{LLL} \prod_{i < j}[(\Omega_{i}^{\uparrow}-\Omega_{j}^{\uparrow}) (\varpi_{i}^{\downarrow}-\varpi_{j}^{\downarrow})^{*}] \mathrm{perm} [G(\Omega_{i}^{\uparrow},\varpi_{j}^{\downarrow})].
\end{equation}
This wavefunction describes pairing of CBs and anti-CBs. This time after the Jain and Kamilla procedure
\begin{equation}
\tilde \Psi_{\textrm{CB}}(\{\Omega^\uparrow\},\{\varpi^\downarrow\})=\prod_{i < j}[ (\Omega^\uparrow_i-\Omega^\uparrow_j) (\varpi^\downarrow_i-\varpi^\downarrow_j)^{*}] \mathrm{perm} [\tilde G(\Omega_{i}^{\uparrow},\varpi_{j}^{\downarrow})]
\end{equation}
with
\begin{equation}
\tilde G(\Omega_{i}^{\uparrow},\varpi_{j}^{\downarrow})=\bigg[ \sum_{n=0}^{N_\textrm{LL}-1} \sum_{m=-(n+\mathcal Q)}^{n+\mathcal Q} g_{n} \tilde{Y}_{\mathcal Q,n, m}\left(\Omega_{i}^{\uparrow}\right) \tilde{Y}^*_{\mathcal Q, n,m}\left(\varpi_{j}^{\downarrow}\right)\bigg].
\end{equation}
Strictly speaking the Jain-Kamilla procedure is not valid for the LLL projection of CB orbitals, however it has been suggested \cite{Balram} that this remains a good approximation to the correct LLL projection procedure even in this case. 

\section{ICCFL}
We can write down the ICCFL state of Ref.~\cite{ICCFL} for a particle number consistent with a filled shell configuration of CF bonding orbitals. For example, for $N_\uparrow=N_\downarrow=6$, we have a total of $N=12$ CFs and hence enough to fill the first three CF shells with $n=0,1,2$. The CFs experience flux $Q=\frac{1}{2}$. The Jain-Kamilla LLL projected wavefunction is then

\begin{equation}
\tilde \Psi_{\textrm{ICCFL}}(\{\Omega^\uparrow\},\{\Omega^\downarrow\})=\prod_{i < j}[ (\Omega^\uparrow_i-\Omega^\uparrow_j)^2 (\Omega^\downarrow_i-\Omega^\downarrow_j)^2] \mathrm{det} [\tilde Y_i(\Omega_{j}^{\uparrow})\dots \tilde Y_i(\Omega_{j}^{\downarrow})]
\end{equation}
where $i=(Q,n,m)$ runs over the $N$ indices of filled CF orbitals and we take the determinant of an $N\times N$ matrix corresponding to filling the $N$ orbitals with CFs from either layer. This wavefunction has no free variational parameters since we have assumed all CFs are in the bonding orbitals in this state, as predicted by mean-field theory\cite{ICCFL}. We also considered the ICCFL wavefunction for the case where we have both bonding and anti-bonding orbitals, however this does not yield significantly improved overlaps.

\section{\uppercase{Supplementary figures}}

In this section we present some further numerical results. In Fig.~\ref{fig:overlaps_ICCFL}a we show the overlap with the ICCFL state of Ref.~\cite{ICCFL}. The ICCFL states appears to capture some of the relevant correlations around $d\sim \ell_B$, however the overlap does not reach order unity anywhere (this wavefunction has no free variational parameters). In Fig.~\ref{fig:overlaps_ICCFL}b we show the overlap with a paired state of composite bosons described in the section above. This trial state of paired CBs performs well at intermediate distances $d\sim\ell_B$, but does not have high overlaps with the 111 state (consistent with the findings of \cite{ShouCheng}) and does not capture the CFL at large distances.

In Figs.~\ref{fig:overlaps_VPs} and \ref{fig:energies_VPs} we show results for different system sizes and different numbers of variational parameters. By increasing the number of variational parameters, we obtain better overlaps and energies at small $d/\ell_B$ since we are including higher momentum orbitals, which in turn allows us to form more tightly bound pairs. Both the $p$-wave CF/CF and the $s$-wave CF/anti-CF wavefunctions perform well at small interlayer separations if sufficiently many variational parameters are included, however in general the $s$-wave trial wavefunction performs better. In Figs.~\ref{fig:overlaps_max_VPs} and \ref{fig:energies_max_VPs} we show the same overlaps and energies for the maximum number of variational parameters employed to show that both trial states capture the 111 state accurately when sufficiently many variational parameters are included.


\begin{figure}[H]
    \centering
    \includegraphics[width=1\columnwidth]{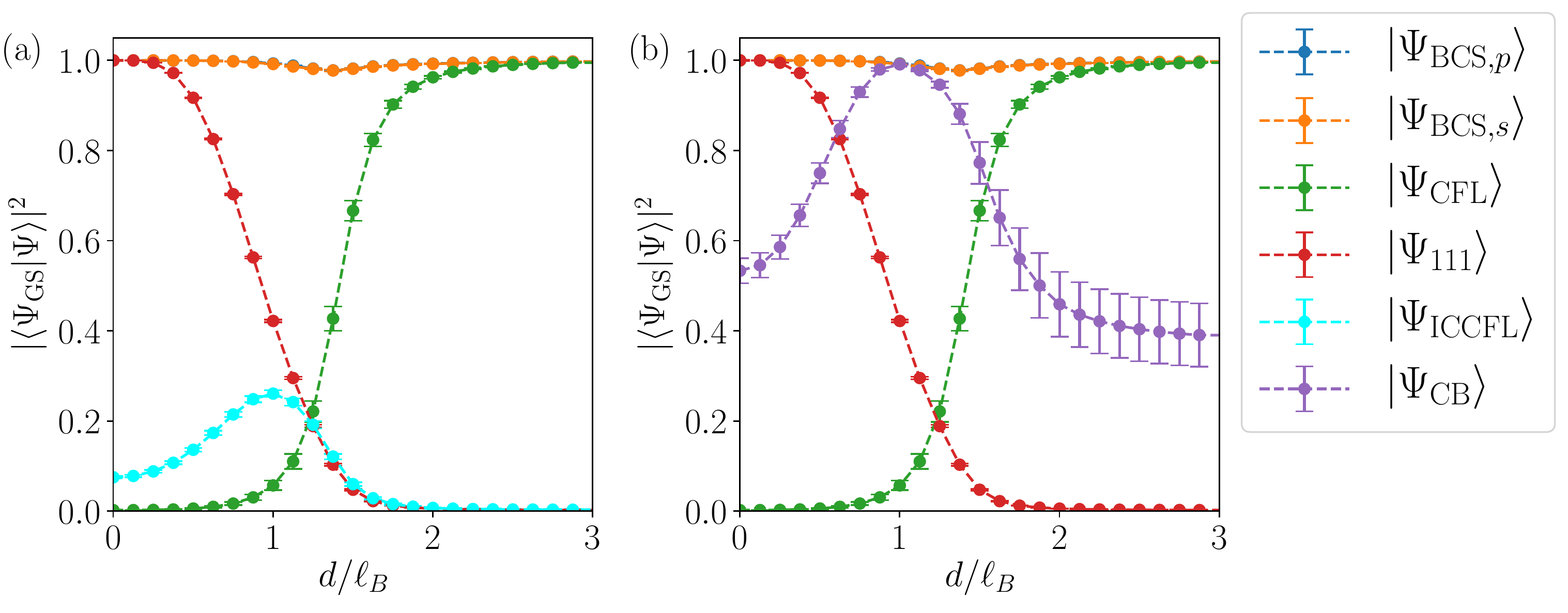}
    \caption{Exact diagonalization results for the overlap of the trial wavefunctions with the true ground state for $6+6$ electrons. We compare the overlap of our $s$-wave BCS state with the previously proposed $p$-wave BCS state of Refs.~\cite{Simon1,Simon2}, the ICCFL state\cite{ICCFL} and a paired state of composite bosons/anti-composite bosons similar to the trial state proposed in Ref.~\cite{ShouCheng}.  In both figures the $s$- and $p$-wave curves are almost exactly on top of each other.  In the ICCFL,  all CFs are in the bonding orbitals, and therefore this wavefunction has no variational parameters. The BCS and CB trial states both have five variational parameters. We also show the overlaps with the CFL state and the 111 state.}
    \label{fig:overlaps_ICCFL}
\end{figure}

\begin{figure}[H]
    \centering
    \includegraphics[width=\columnwidth]{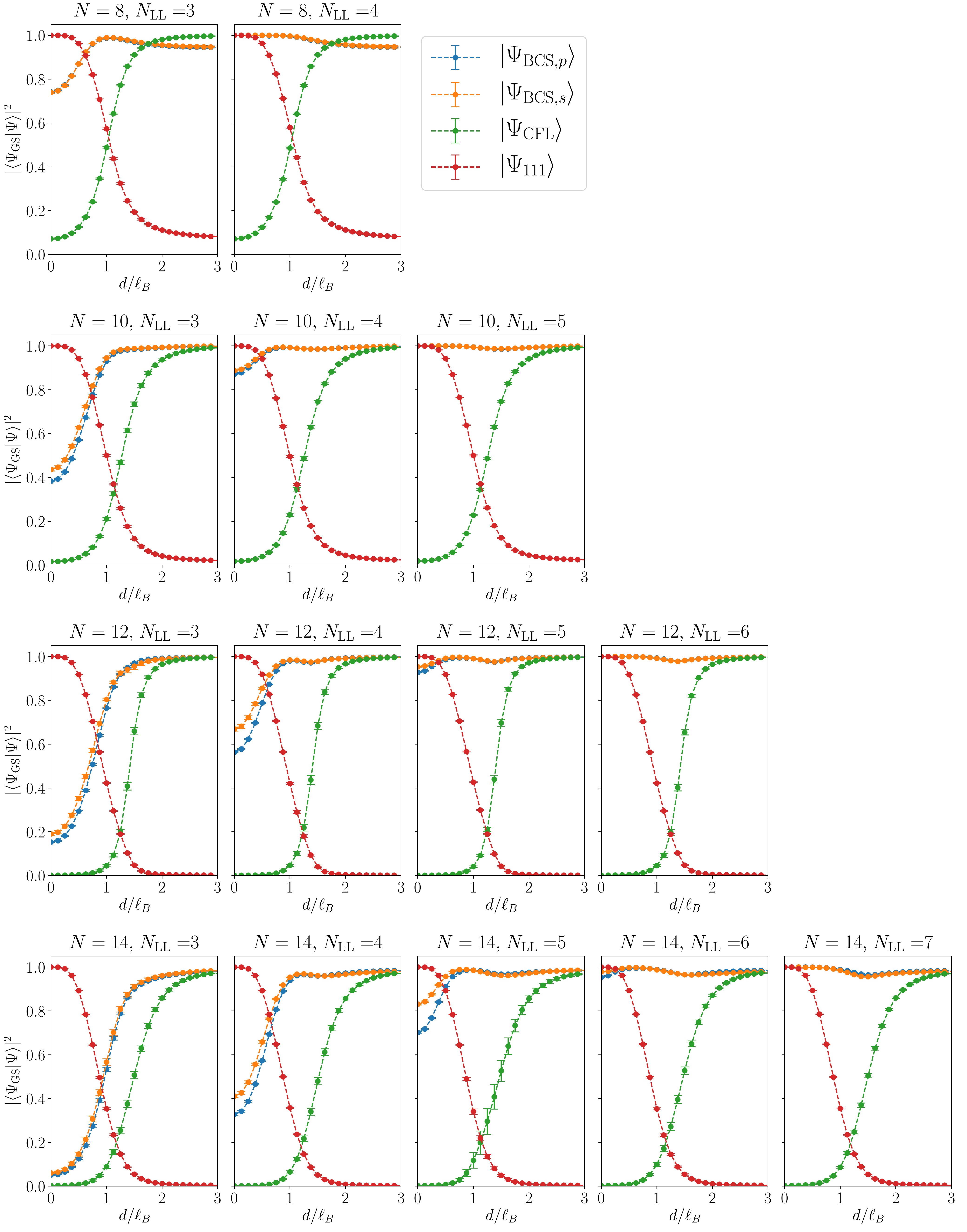}
    \caption{Exact diagonalization results for the overlap of the trial wavefunctions with the true ground state  for the balanced layer case, for different system sizes and different numbers of variational parameters. We include variational parameters $g_n$ for $n=0,\dots,N_\textrm{LL}-1$, such that the number of variational parameters is $N_\textrm{LL}-1$ (since an overall rescaling of all $g_n$ results in the same trial wavefunction after normalization).  We compare the overlap of our $s$-wave BCS state with the previously proposed $p$-wave BCS state of Refs.~\cite{Simon1,Simon2}. The $s$-wave and $p$-wave states are taken to have the same number of variational parameters.  In some of the plots furthest to the right, the $s$- and $p$-wave curves are almost exactly on top of each other. We also show the overlaps with the CFL state and the 111 state. By CFL we denote either a CF Fermi sea (for $N=12$) or a Hund's rule state (for other $N$). Both BCS states can reproduce the CFL at large $d$, except when there is more than one CF in a partially filled shell (this happens for the $4+4$ system size for example).}
    \label{fig:overlaps_VPs}
\end{figure}

\begin{figure}[H]
    \centering
    \includegraphics[width=\columnwidth]{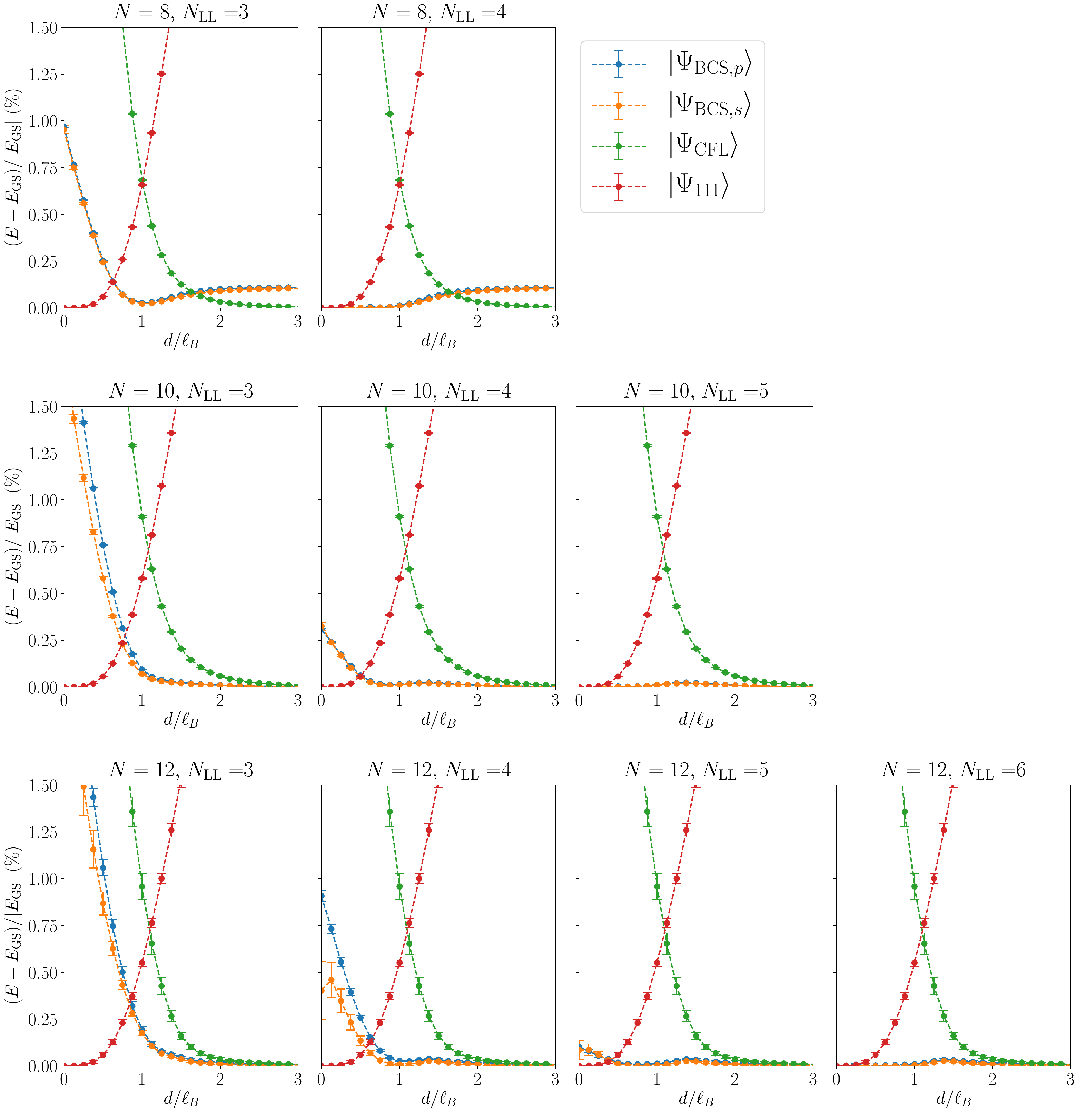}
    \caption{Exact diagonalization results for the energy of the trial wavefunctions compared with the true ground state energy for different system sizes and different numbers of variational parameters. We compute the energy of the trial state which has the best overlap with the ED ground state. We also compute the energy of the 111 and the CFL states.}
    \label{fig:energies_VPs}
\end{figure}

\begin{figure}[H]
    \centering
    \includegraphics[width=\columnwidth]{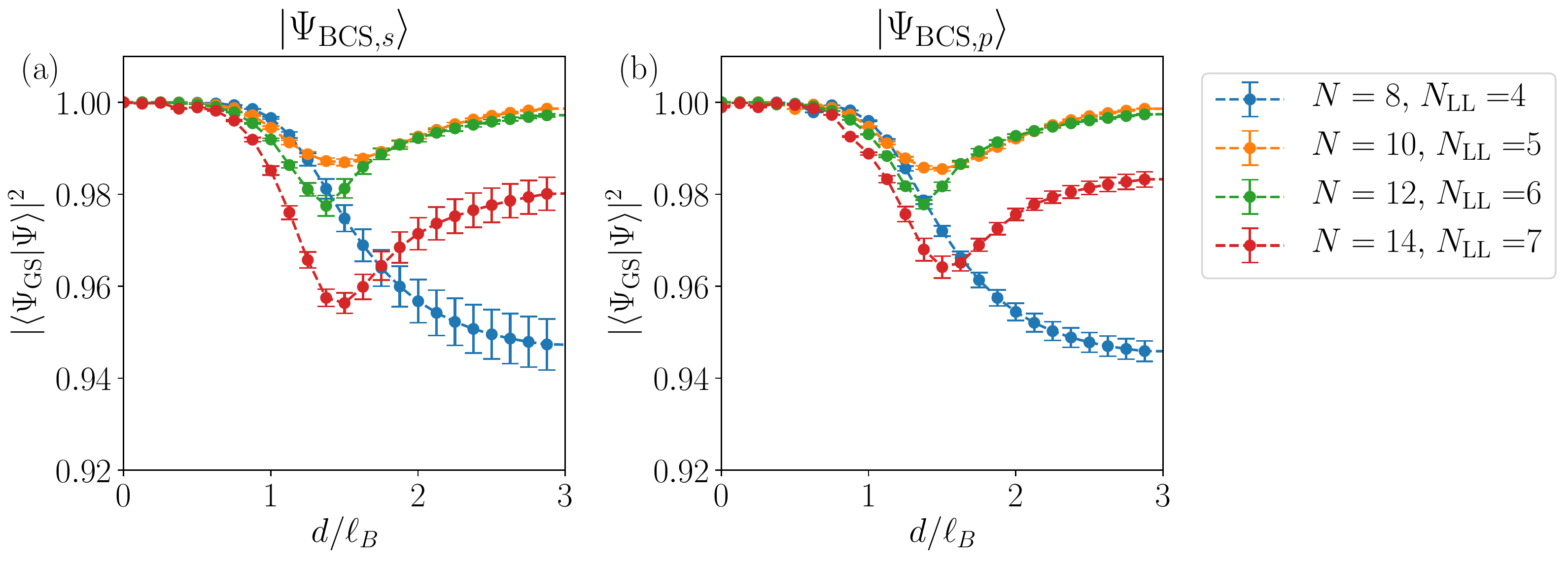}
    \caption{Same data as Fig.~\ref{fig:overlaps_VPs}, but only for the largest number of variational parameters employed for a given $N$.}
    \label{fig:overlaps_max_VPs}
\end{figure}

\begin{figure}[H]
    \centering
    \includegraphics[width=\columnwidth]{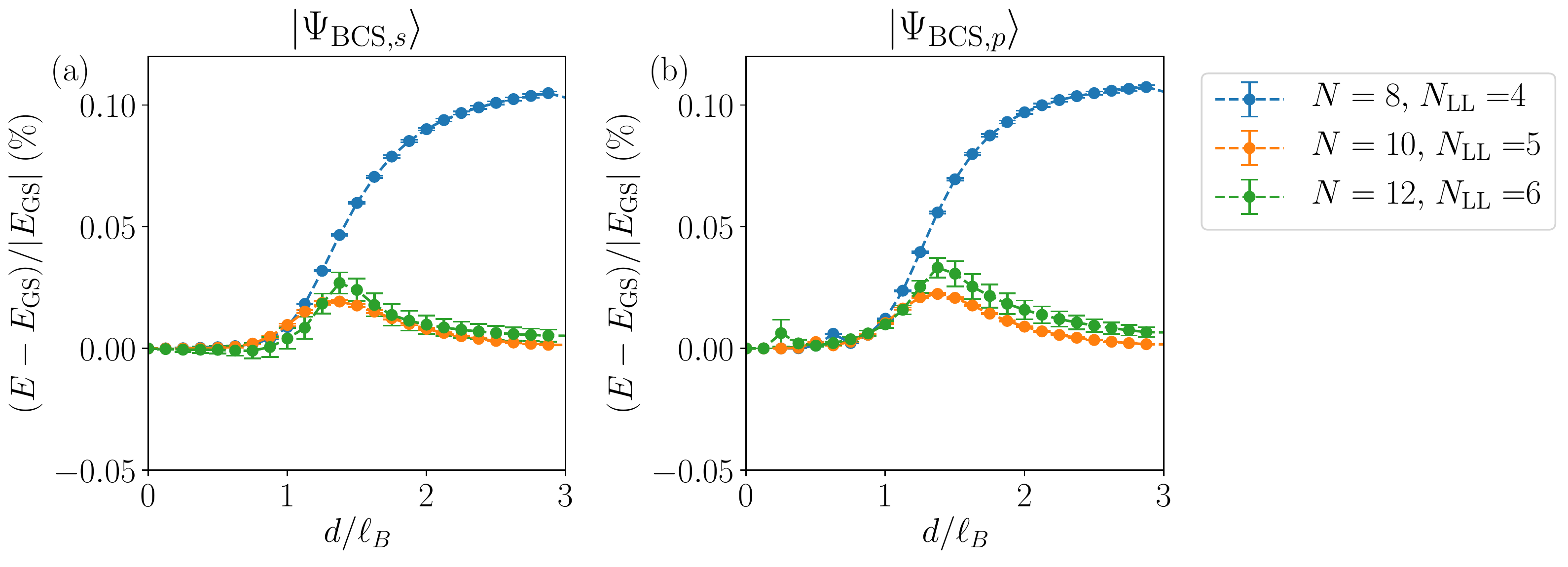}
    \caption{Same data as Fig.~\ref{fig:energies_VPs}, but only for the largest number of variational parameters employed for a given $N$.}
    \label{fig:energies_max_VPs}
\end{figure}

\section{\uppercase{BCS parameters}}
If we define the CF orbitals including the Jastrow factors as
\begin{equation}
    \mathcal Y_{Q,n,m}(\Omega_{i})=J_i\tilde{ Y}_{Q,n, m}\left(\Omega_{i}\right),
\end{equation}
then we can write the $s$-wave trial wavefunction as
\begin{equation}
\tilde \Psi_{\textrm{BCS},s}(\{\Omega^\uparrow\},\{\varpi^\downarrow\})=   \mathrm{det} \bigg[ \sum_{n=0}^{N_\textrm{LL}-1} \sum_{m=-(n+Q)}^{n+Q} g_{n} \mathcal Y_{Q,n,m}(\Omega_{i}^{\uparrow})\mathcal Y_{Q,n,m}^*(\varpi_{j}^{\downarrow}) \bigg].
\end{equation}
One subtlety arises regarding the normalization of the CF orbitals $\mathcal Y_{Q,n,m}(\Omega_i)$, since these are no longer single-particle wavefunctions. We need to extract the relative normalization of the orbitals with different $n$. To do so, for example for $N_\uparrow=N_\downarrow=4$ we write down the CFL wavefunctions filling the four orbitals in each layer with $n$ given by $(0,0,1,1)$, $(0,0,2,2)$ and $(1,1,2,2)$. We compute the normalization of the CF orbitals $\mathcal Y_{Q,n,m}(\Omega_i)$ by computing the ratio of the norm of these many-body CFL wavefunction. 

We can translate from angular momentum $l=Q+n$ to linear momentum $k=l/R$, where $R=\sqrt{q}\ell_B$ is the radius of the sphere. Recall the second quantized BCS wavefunction
\begin{equation}
    |\Psi_\textrm{BCS}\rangle=\prod_{\mathbf{k}}\left(1+g_{\mathbf{k}} c_{\mathbf{k}, \uparrow}^{\dagger} d_{\mathbf{k}, \downarrow}^{\dagger}\right)|0\rangle,
    \label{eq:BCS}
\end{equation}
where $|0\rangle$ is the state with the upper layer empty and the LLL of the lower layer filled. Eq.~\eqref{eq:BCS} can equally be interpreted as a BEC wavefunction, by re-writing it as 
\begin{equation}
    |\Psi_\textrm{BEC}\rangle=e^{\sum_{\mathbf{k}}g_{\mathbf{k}} c_{\mathbf{k}, \uparrow}^{\dagger} d_{\mathbf{k}, \downarrow}^{\dagger}}|0\rangle.
\end{equation}
This motivates us to consider a smooth crossover between a BCS-like regime at large interlayer separation and a BEC-like regime at small interlayer separation\cite{Crossover}. Note that we need to project Eq.~\eqref{eq:BCS} to a definite $S_z$ sector in order to be able to write down the real-space representation. Let us focus on the balanced case, i.e.~we sum over the possible sets $\{g_\mathbf{k}\}$ of orbitals to fill:
\begin{equation}
    |\Psi_\textrm{BCS}\rangle=\sum_{\{g_\mathbf{k}\}}\prod_{\mathbf{k}}\left(g_{\mathbf{k}} c_{\mathbf{k}, \uparrow}^{\dagger} d_{\mathbf{k}, \downarrow}^{\dagger}\right)|0\rangle,
\end{equation}
where $|\{g_\mathbf{k}\}|=N_\uparrow$. And now we can calculate the occupation of each orbital according to
\begin{equation}
    n_\mathbf{k}=\langle c_{\mathbf{k}, \uparrow}^{\dagger}c_{\mathbf{k}, \uparrow}\rangle=\langle d_{\mathbf{k}, \downarrow}^{\dagger}d_{\mathbf{k}, \downarrow}\rangle=\frac{\sum_{\{g_\mathbf{q}\}\ni g_\mathbf{k} }\prod_{\mathbf{q}}(g_{\mathbf{q}})^2}{\sum_{\{g_\mathbf{q}\}}\prod_{\mathbf{q}}(g_{\mathbf{q}})^2}.
    \label{eq:n_k_ED}
\end{equation}
For an $s$-wave superconductor, BCS theory predicts
\begin{equation}
    n_\mathbf{k}=\frac{1}{2}\bigg(1-\frac{\varepsilon_\mathbf{k}}{\sqrt{\varepsilon_\mathbf{k}^2+\Delta^2}}\bigg),
    \label{eq:n_k}
\end{equation}
where $\varepsilon_\mathbf{k}=\frac{k^2}{2m}-E_F=\frac{k^2-k_F^2}{2m}$. The typical size of a Cooper pair is set by the coherence length 
\begin{equation}
    \xi=\frac{v_F}{\Delta}=\frac{k_F/m}{\Delta}.
\end{equation}
We show results for $n_\mathbf{k}$ in Fig.~\ref{fig:n_k}, which we can interpret as evidence of a BEC-BCS crossover. At $d/\ell_B\ll1$, $n_k$ is a smeared out step function and the order parameter satisfies $\Delta\gtrsim E_F$, this is the BEC limit. One the other hand, at $d/\ell_B\gg1$, $n_\mathbf{k}$ is a sharp step function and the order parameter satisfies $\Delta\ll E_F$. This is the BCS limit.  

We note that the formulas for $n_{\mathbf{k}}$ and $\xi$ depend essentially on the ratio $\Delta/E_F $ and do not require any assumption about the value of the CF effective mass. The values of  $n_{\mathbf{k}}$ and $\xi$ plotted in Fig. 1b of the main text were obtained by fitting the BCS form to our results for $n_{\mathbf{k}}$, as shown in Fig. S7, below.

\begin{figure}[H]
    \centering
    \includegraphics[width=\columnwidth]{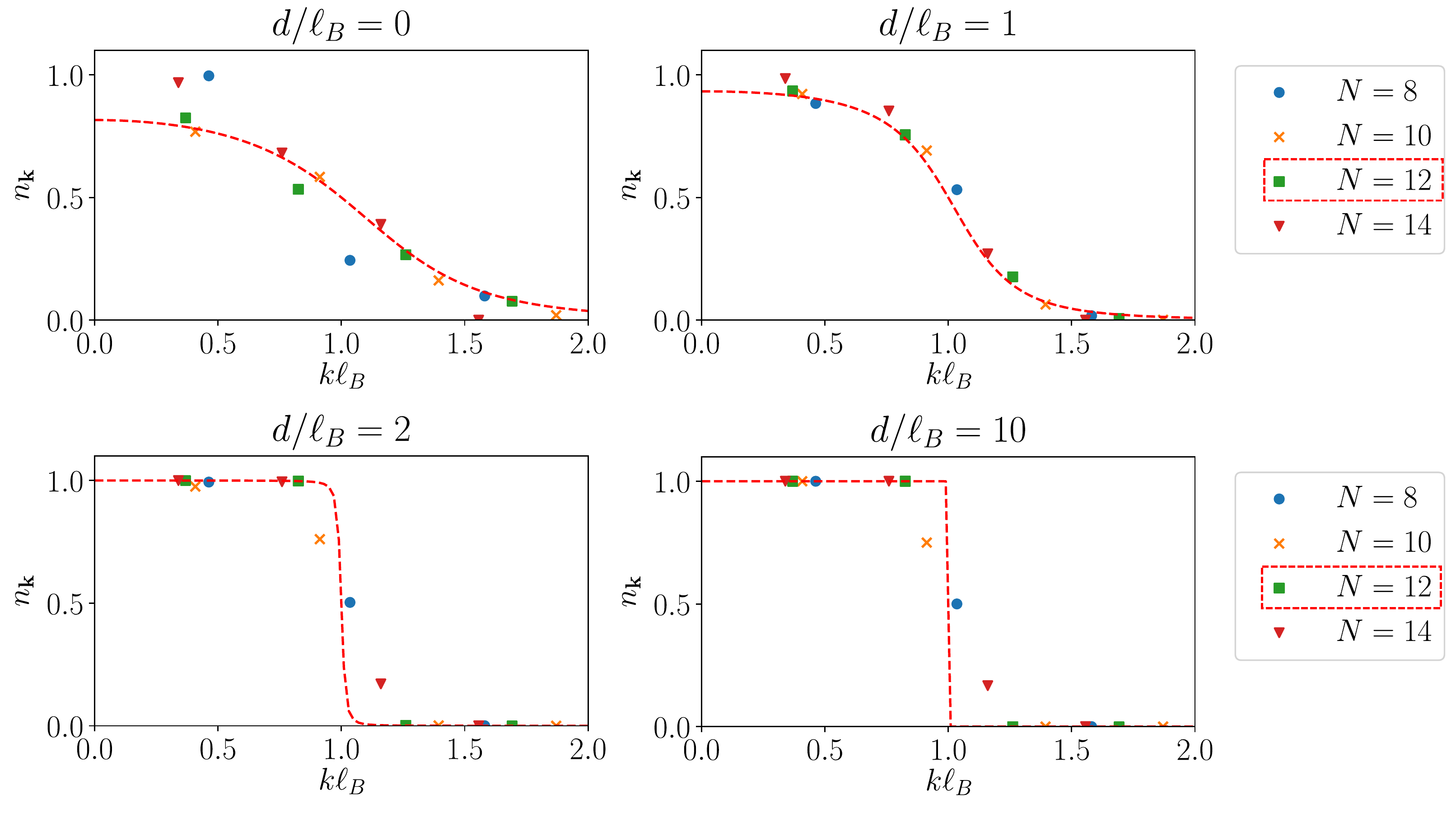}
    \caption{ Values of the BCS parameters $n_\mathbf{k}$ for the best $s$-wave trial state at four selected interlayer separations $d$. At large $d$ we have a sharp Fermi surface, whereas at small $d$ we have significant pairing in a large momentum range, which corresponds to tightly bound CF/anti-CF pairs. The dashed line shows the best fit of the BCS form Eq.~\eqref{eq:n_k}. For the fit we only use the $N=12$ data, since this is a closed shell configuration at large $d/\ell_B$. For all $N$ we use 3 variational parameters (i.e.~$N_\textrm{LL}=4)$.}
    \label{fig:n_k}
\end{figure}

\section{Monte-Carlo procedure}
We follow the approach outlined in Ref.~\cite{Kamilla_thesis}. For the calculation of the overlaps we use the probability distribution of the 111 state as a sampling distribution for the Monte-Carlo (MC) algorithm. This choice guarantees that we are sampling from the part of the Hilbert space in which the ground state lies for small $d/\ell_B$. For large $d/\ell_B$, we find that the MC errors lie within an acceptable range, even for this choice of sampling. 

For the overlaps of the ED ground state with the 111 state, the CFL and the $p$-wave BCS state, we work in terms of electron coordinates in both layers. We start with a random distribution of the electron coordinates $(\{\Omega^\uparrow\},\{\Omega^\downarrow\})$ and perform a random walk on the sphere according to \cite{Kamilla_thesis}. At each MC step, we propose a move of one particle to a new position and accept or reject the move based on the Metropolis algorithm for the probability distribution
\begin{equation}
    \rho(\{\Omega^\uparrow\},\{\Omega^\downarrow\})=|\Psi_{111}(\{\Omega^\uparrow\},\{\Omega^\downarrow\})|^2.
\end{equation}
After a suitable equilibration period, the electron coordinates will be distributed according to the probability distribution $\rho(\{\Omega^\uparrow\},\{\Omega^\downarrow\})$ and we can generate MC samples $(\{\Omega^\uparrow\}_I,\{\Omega^\downarrow\}_I)$ where $I=0,\dots,N_\textrm{MC}-1$. The number of MC samples $N_\textrm{MC}$ was chosen up to $10^8$ depending on system size.
Then, for example, the overlap of the ED ground state with the CFL will be
\begin{align}
    \langle\Psi_\textrm{GS}|\Psi_\textrm{CFL}\rangle&=\int \diff\Omega_1^\uparrow\dots \diff\Omega_{N_\uparrow}^\uparrow\diff\Omega_1^\downarrow\dots \diff\Omega_{N_\downarrow}^\downarrow \Psi_\textrm{GS}^*(\{\Omega^\uparrow\},\{\Omega^\downarrow\})\Psi_\textrm{CFL}(\{\Omega^\uparrow\},\{\Omega^\downarrow\})\\
    &=\frac{1}{N_\textrm{MC}}\sum_{I=0}^{N_\textrm{MC}-1}\frac{\Psi_\textrm{GS}^*(\{\Omega^\uparrow\}_I,\{\Omega^\downarrow\}_I)\Psi_\textrm{CFL}(\{\Omega^\uparrow\}_I,\{\Omega^\downarrow\}_I)}{\rho(\{\Omega^\uparrow\}_I,\{\Omega^\downarrow\}_I)}j(\{\Omega^\uparrow\}_I,\{\Omega^\downarrow\}_I)
\end{align}
where
\begin{equation}
    j(\{\Omega^\uparrow\},\{\Omega^\downarrow\})=\prod_{n=1}^{N_\uparrow}\sin\theta_n^\uparrow\prod_{n=1}^{N_\downarrow}\sin\theta_n^\downarrow
\end{equation}
is the Jacobian for the area element on the sphere. We proceed similarly for the overlap with the $s$-wave trial state, except that we work in terms of electron coordinates in the top layer and hole coordinates in the bottom layer, i.e.~$(\{\Omega^\uparrow\},\{\varpi^\downarrow\})$ and use the probability distribution
\begin{equation}
    \rho(\{\Omega^\uparrow\},\{\varpi^\downarrow\})=|\Psi_{111}(\{\Omega^\uparrow\},\{\varpi^\downarrow\})|^2.
\end{equation}

\end{widetext}

\end{document}